\documentclass[10pt,a4paper,notitlepage,twocolumn,showpacs,preprintnumbers,superscriptaddress,nofootinbib]{revtex4-1}
\usepackage{float}
\usepackage{amsmath,amssymb}
\usepackage{cancel}
\usepackage{graphicx}

\usepackage{bm}
\usepackage{tensor}
\usepackage{comment}
\usepackage{ifpdf}
\usepackage{slashed}
\usepackage{color}
\usepackage[mathscr]{eucal}

\ifpdf
  \usepackage{graphicx}     
  \usepackage[bookmarksopen,colorlinks=true,linkcolor=bblue,citecolor=bblue,urlcolor=ppink]{hyperref}
\else     
  \usepackage[dvipdfmx]{}     
  \usepackage[dvipdfmx,bookmarksopen,colorlinks=true,linkcolor=bblue,citecolor=ppink,urlcolor=darkred]{hyperref}
\fi

\definecolor{red}{rgb}{1,0,0}
\definecolor{darkred}{rgb}{0.6,0,0}
\definecolor{darkgreen}{rgb}{0.992447,0.623778,0.034597}
\definecolor{ppink}{rgb}{1,0.4,0.4}
\definecolor{bblue}{rgb}{0.284602,0.317763,0.963947}
\begin{document}


\def\o{\over}
\def\beq{\begin{eqnarray}}
\def\eeq{\end{eqnarray}}
\newcommand{\cc}{ {\rm c.c.} }
\newcommand{\gsim}{ \mathop{}_{\textstyle \sim}^{\textstyle >} }
\newcommand{\lsim}{ \mathop{}_{\textstyle \sim}^{\textstyle <} }
\newcommand{\vev}[1]{ \left\langle {#1} \right\rangle }
\newcommand{\bra}[1]{ \langle {#1} | }
\newcommand{\ket}[1]{ | {#1} \rangle }
\newcommand{\EV}{ {\rm eV} }
\newcommand{\KEV}{ {\rm keV} }
\newcommand{\MEV}{ {\rm MeV} }
\newcommand{\GEV}{ {\rm GeV} }
\newcommand{\TEV}{ {\rm TeV} }
\newcommand{\1}{\mbox{1}\hspace{-0.25em}\mbox{l}}
\newcommand{\headline}[1]{\noindent{\bf #1}}
\def\diag{\mathop{\rm diag}\nolimits}
\def\Spin{\mathop{\rm Spin}}
\def\SO{\mathop{\rm SO}}
\def\O{\mathop{\rm O}}
\def\SU{\mathop{\rm SU}}
\def\U{\mathop{\rm U}}
\def\Sp{\mathop{\rm Sp}}
\def\SL{\mathop{\rm SL}}
\def\tr{\mathop{\rm tr}}
\def\mpl{M_{PL}}

\def\Mpl{M_{\rm Pl}}
\def\GeV{{\rm GeV}}
\newcommand{\Red}[1]{\textcolor{red}{#1}}
\newcommand{\Blue}[1]{\textcolor{blue}{\bfseries\sffamily #1}}


\title{More about Q-ball with elliptical orbit}

\author{Fuminori Hasegawa}
\email[e-mail: ]{fuminori@icrr.u-tokyo.ac.jp}
\affiliation{ICRR, University of Tokyo, Kashiwa, Chiba 277-8582, Japan}
\affiliation{Kavli IPMU (WPI), UTIAS, University of Tokyo, Kashiwa, Chiba 277-8583, Japan}
\author{Jeong-Pyong Hong}
\email[e-mail: ]{hjp0731@snu.ac.kr}
\affiliation{Center for Theoretical Physics, Department of Physics and Astronomy, Seoul National University, Seoul 08826, Korea}
\author{Motoo Suzuki}
\email[e-mail: ]{m0t@icrr.u-tokyo.ac.jp}
\affiliation{ICRR, University of Tokyo, Kashiwa, Chiba 277-8582, Japan}
\affiliation{Kavli IPMU (WPI), UTIAS, University of Tokyo, Kashiwa, Chiba 277-8583, Japan}
\begin{abstract}
Q-balls formed from the Affleck-Dine field have rich cosmological implications and have been extensively studied from both theoretical and simulational approaches. From the theoretical point of view, the exact solution of the Q-ball was obtained and it shows a circular orbit in the complex plane of the field value.
In practice, however, it is reported that the Q-ball that appears after the Affleck-Dine mechanism has an {\it elliptical} orbit, which carries larger energy per unit $U(1)$ charge than the well-known solution with a circular orbit. 
We call them ``elliptical" Q-balls.
In this paper, we report the first detailed investigation of the properties of the elliptical Q-balls by $3$D lattice simulation. The simulation results indicate that the elliptical Q-ball has an almost spherical spatial profile with no nodes, and we observed a highly elliptic orbit that cannot be described through small perturbations around the ground state Q-ball. Higher ellipticity leads to more excitation of the energy, whose relation is also derived as a dispersion relation. Finally, we derive two types of approximate solutions by extending the Gaussian approximation and considering the time-averaged equation of motion and we also show the consistency with the simulation results.
\end{abstract}

\date{\today}
\maketitle
\preprint{IPMU 19-0043}
\section{Introduction} 

Q-ball is a non-topological soliton in a complex scalar field theory with a global $U(1)$ symmetry~\cite{Coleman:1985ki}. The solution corresponds to the spherically symmetric field configuration minimizing the energy of the system with the fixed $U(1)$ charge. Such a solution exists for the potential $V(\phi)$ of the complex scalar field $\phi$ if $V(\phi)/|\phi|^2$ has the minimum at $\phi\neq 0$. One of the characteristics of the solution is that the configuration of $\phi$ has time dependence and the orbit of $\phi$ in the complex plane is a true circle.  

The formation of the Q-ball has been studied analytically and numerically in some extensions of the standard model. For example, in the supersymmetric extension of the standard model, some scalar condensates with the baryon number form Q-balls after Affleck-Dine baryogenesis~\cite{Affleck:1984fy,Cohen:1986ct,Dine:1995kz,Dvali:1997qv,Kusenko:1997si,Enqvist:1997si, Enqvist:1998xd,Enqvist:1998en,Kasuya:1999wu,Kasuya:2000wx,Enqvist:2000cq,Kasuya:2001hg,Hiramatsu:2010dx}. There, the initial configuration of the complex scalar field is set to be the coherently oscillating state in the complex plane. Due to the instability of such a state, the initial fluctuation of the scalar field grows and eventually develops into the Q-balls.

One non-trivial result from the numerical analysis is the discovery of excited states of Q-balls~\cite{Enqvist:1999mv,Hiramatsu:2010dx,Lozanov:2014zfa}.%
\footnote{In Ref.~\cite{Enqvist:1999mv}, the excited state is named ``Q-axitons''.} 
The excited state has larger energy compared to the energy-minimizing solution with the same charge. 
The configuration of the complex scalar in the excited state draws an elliptical orbit, while the orbit of a genuine Q-balls is a true circle. We call those excited Q-balls as the ``elliptical'' Q-balls and genuine ones as the ground state Q-balls. It is numerically shown that the elliptical Q-balls are semi-stable at least~\cite{Hiramatsu:2010dx} and  they are considered as the ``transients'', which appear during the formation of the circular Q-balls.

Even though the formation of the elliptical Q-balls was pointed out, their properties were not pursued sufficiently. For instance, in the numerical simulation, there is no detailed investigation on how the energy excitation depends on the ellipticity. The theoretical explanation of the elliptical Q-ball solution is also not present.

 It should be noted that the elliptical Q-ball is different from the excited state solution that is discussed in $e.g.$ Refs.~\cite{Volkov:2002aj,Polyakov:2018zvc}. The radial excitation with nodes~\cite{Volkov:2002aj,Polyakov:2018zvc} and the (non-spherical) angular excitation~\cite{Volkov:2002aj} of the Q- ball solution are studied as possible forms of the excitation. However, the Q-balls produced in the lattice simulation seem to have no nodes and are almost spherically symmetric (See Sec. III for more details.). Furthermore, perturbations around the ground state Q-ball~\cite{Coleman:1985ki,Lee:1991ax,Smolyakov:2017axd} do not explain the dispersion relation of the elliptical Q- ball.

In this paper, we investigate the detailed properties of the elliptical Q-ball for a scalar potential motivated by the AD mechanism in gravity mediation where the properties of the ground Q-ball size and energy are given by the simple formulas.
We perform 3D lattice simulations and present the spatial profiles and the dispersion relation of the elliptical Q-ball. Then, we will derive some approximate solutions of the elliptical Q-ball, which we will show are in rough agreement with the lattice simulation results.
We also discuss the stability of the elliptical Q-ball by considering perturbations to the approximate solution.

The paper is organized as follows. In Sec.~\ref{sec:2}, we review the ground state Q-ball solution and its properties in gravity mediation model. In Sec.~\ref{sec:3}, we describe the setup of our numerical simulation. In Sec.~\ref{sec:4}, we show the numerical results on the formation of the elliptical Q-balls and their detailed properties. In Sec.~\ref{sec:5}, we derive the approximate elliptical Q-ball solutions and discuss the consistency with the simulations. The final section is devoted to our conclusion.

\section{Ground State Q-ball}
\label{sec:2}

In this section, let us review the ground state Q-ball solution. In particular, we focus on the Q-ball in the gravity mediation model.

Let us consider the theory of a complex scalar field $\phi$ with the Lagrangian density,
\begin{align}
\label{eq:lagrangian}
\mathcal{L}=\partial_\mu\phi\partial^\mu\phi^*-V(\phi).
\end{align}
Here, $V(\phi)$ is a potential with the $U(1)$ symmetry, which is invariant under the phase rotation of $\phi$. The Noether charge of this $U(1)$ symmetry is given by
\begin{align}
\label{eq:charge}
Q=\frac{1}{i}\int d^3x(\phi^*\dot{\phi}-\phi\dot{\phi}^*).
\end{align}
The energy is represented as
\begin{align}\label{ene}
E=\int d^3x\left[\dot\phi\dot\phi^*+\nabla\phi\nabla\phi^*+V(\phi)\right].
\end{align}

To figure out the ground state Q-ball solution, let us consider the energy-minimizing condition for a fixed charge. Using the method of Lagrange multipliers, one must find an extremum of 
\begin{align}\label{omeene}
E_{\omega}=&\int d^3x\left[\dot\phi\dot\phi^*+\nabla\phi\nabla\phi^*+V(\phi)\right] \nonumber\\
+&\omega\left(Q-\int d^3x~i(\dot\phi^*\phi-\phi^*\dot\phi)\right)\\
=&\int d^3x\left[|\dot\phi-i\omega \phi|^2+|\nabla\phi|^2-\omega^2|\phi|^2+V(\phi)\right]+\omega Q,
\end{align}
where  $\omega$ is a Lagrange multiplier.
From the first term of the integrand in the second line, the time-dependent part of $\phi$ is minimized for the choice
\begin{align}\label{sepa}
\phi({\bf x},t)=\frac{1}{\sqrt{2}}\Phi({\bf x})e^{i\omega t}.
\end{align}
Note that this formula indicates that the orbit in the complex plane is a true circle. Then, the  Eq.~(\ref{omeene}) is reduced to 
\begin{align}\label{omeene2}
E_{\omega}=&\int d^3x\left[\frac12(\nabla\Phi)^2-\frac12\omega^2\Phi^2+V(\Phi)\right]+\omega Q.
\end{align}
As the extremum condition, we then obtain the equation of motion for $\Phi({\bf x})$,
\begin{align}
\label{eomQ}
\frac{d^2\Phi}{dr^2}+\frac{2}{r}\frac{d\Phi}{dr}+\omega^2\Phi-\frac{\partial V(\Phi)}{\partial \Phi}=0,
\end{align}
where $r$ is the radial coordinate and we assume that $\Phi$ is spherically symmetric in the position space. From the finiteness of the energy and the continuity of the solution, we impose the boundary condition,
\begin{align}
\Phi'(0)=0,~~\Phi(\infty)=0.
\end{align}
The condition~\cite{Coleman:1985ki} for the existence of the solution is given by
\begin{align}
& ^\exists \omega~~\text{which satisfies} \nonumber\\ 
&~~{\rm Min}\left[\frac{2V(\Phi)}{\Phi^2}\right]<\omega^2<\left.\frac{2V(\Phi)}{\Phi^2}\right|_{\Phi=0}\,.
\end{align}
This condition claims that $V(\Phi)$ must be ``shallower" than the quadratic.

In the following, the potential is specified to the gravity mediation type,
\begin{align}
V(\phi)=m^2|\phi|^2\left[1+K\ln\left(\frac{|\phi|^2}{M_*^2}\right)\right],
\end{align}
where $M_*$ is the renormalization scale defining the mass parameter, $m$, of the complex scalar field. The constant $K$ is for example determined as  $K=-0.01\sim-0.1$ by the spectrum of the minimal supersymmetric standard model. Throughout this paper, we take,
\begin{align}
K=-0.1.
\end{align}
We have confirmed that the qualitative results in the following are the same for $K=-0.01$.

In gravity mediation scenarios, the equation of motion is numerically solved and it is shown that the solution is well approximated by the Gaussian spatial profile~\cite{Enqvist:1998en},
\begin{align}
\Phi(r)=\Phi(0)e^{-r^2/R^2},
\end{align}
where $R$ is the constant which corresponds to the radius of the Q-ball.
By substituting this into the Eq.~(\ref{eomQ}), one obtains the following relation,
\begin{align}
\omega&\simeq m\left[1+2|K|-|K|\ln\left(\frac{\Phi(0)^2}{2M_*^2}\right)\right]^{1/2}\sim m,\\
\label{eq:R}
R&\simeq \left(\frac{2}{|K|}\right)^{1/2}m^{-1}.
\end{align}
Then, the charge and the energy is obtained as 
\begin{align}
Q&\simeq \left(\frac{\pi}{2}\right)^{3/2}m\Phi(0)^{2}R^3,\\
E&\simeq mQ.
\end{align}
We stress that one important feature of the ground state Q-ball in gravity mediation is the dispersion relation, $E\simeq mQ$.
As we will see later, the Q-balls formed in the realistic situation have elliptical orbits and no longer follow this relation.

Hereafter, we redefine the parameters as
\begin{align}
\label{eq:norma}
mt&\rightarrow t,  \nonumber\\ 
m{\bf x}&\rightarrow {\bf x},  \nonumber
\\
mR&\rightarrow R,  \nonumber
\\
\phi/M_*&\rightarrow\phi,\\
\omega/m&\rightarrow \omega,\nonumber
\\
Em/M_*^2&\rightarrow E,\nonumber
\\ 
Qm^2/M_*^2&\rightarrow Q\nonumber
, 
\end{align}
for later convenience.

\section{Simulation Setup}
\label{sec:3}

In this section, we perform simulations in the gravity mediation model and show the formation and properties of the elliptical Q-balls. While there is no rigid theory as in the previous section on the elliptical Q-ball solution, we will give the approximate theory in the next section, which turns out to roughly agree with the simulation results.

In our analysis, we use the package PyCOOL~\cite{Sainio:2012mw}, which we modified for our analysis. The PyCOOL is an object-oriented Python program that uses GPU(s) for simulating the scalar field dynamics in the early universe and performs the 3D simulations.

Following Ref.~\cite{Sainio:2012mw}, we use $N_{\rm grid}=128^3$, $L=8$, $t_i=100$, $t_f=5000$, and $\Delta\tau=0.005$.%
\footnote{For the parameters before the redefinition in Eq.~(\ref{eq:norma}), $N_{\rm grid}=128^3$, $Lm=8$, $t_i=100m^{-1}$, $t_f=5000m^{-1}$, and $m\Delta\tau=0.005$.}
 There, $t_i,~t_f$, and $\tau$ are the initial and final, and the conformal time, respectively.

We assumed the matter-domination after the inflation\footnote{We also performed the simulation in the Minkowski case, but identifying the profile and the elliptical behavior was difficult due to the contamination by undamping higher modes.} and set the initial scale factor unity:~$a(t=t_i)=1$. As the initial condition for the homogeneous AD field, we take 
\begin{align}
\phi(t=t_i)&=1,\\
\frac{d\phi}{dt}(t=t_i)&=i\epsilon_{\rm ini},
\end{align}
where $\epsilon_{\rm ini}$ represents the initial ellipticity of the orbit in the complex plane. We also give the initial Gaussian fluctuation, though the eventual phenomena are known to be irrelevant to this initial condition on the fluctuation, except for the formation time of the objects. In the next section, we will show the result of our simulation where the elliptical Q-balls are formed.

\section{Numerical Results}
\label{sec:4}

\begin{figure}[t]
   \centering
   \includegraphics[width=70mm]{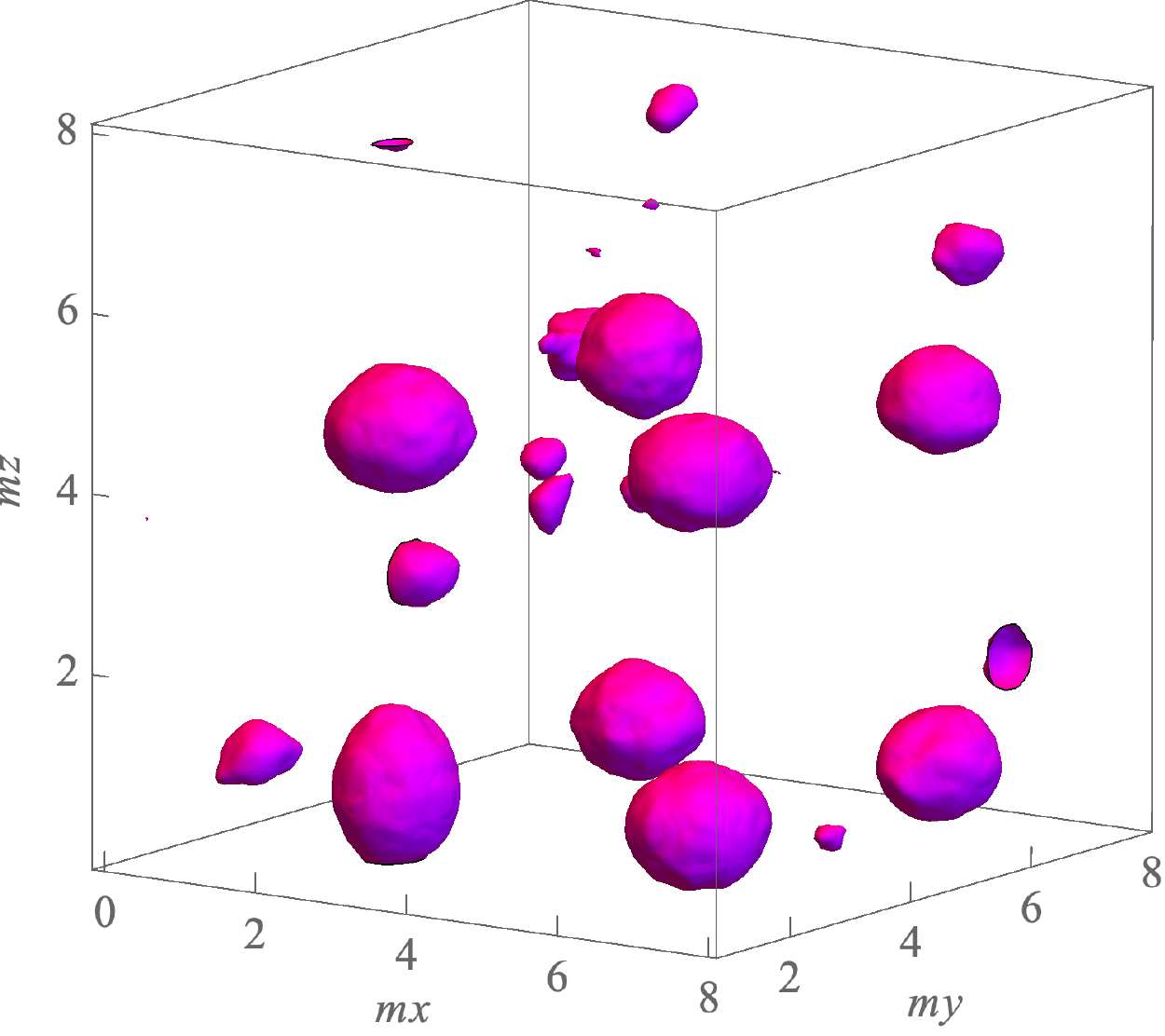}
   \caption{\sl \small Formation of excited Q-balls in 3D lattice simulation at $t=t_f$. We plotted an isosurface of the charge density for the illustration. We set $\epsilon_{\rm ini}=0.2$.} 
   \label{fig:3dp}
\end{figure}

\begin{figure}[ht]
      \begin{center}
  \begin{minipage}{.45\linewidth}
  \includegraphics[width=40mm]{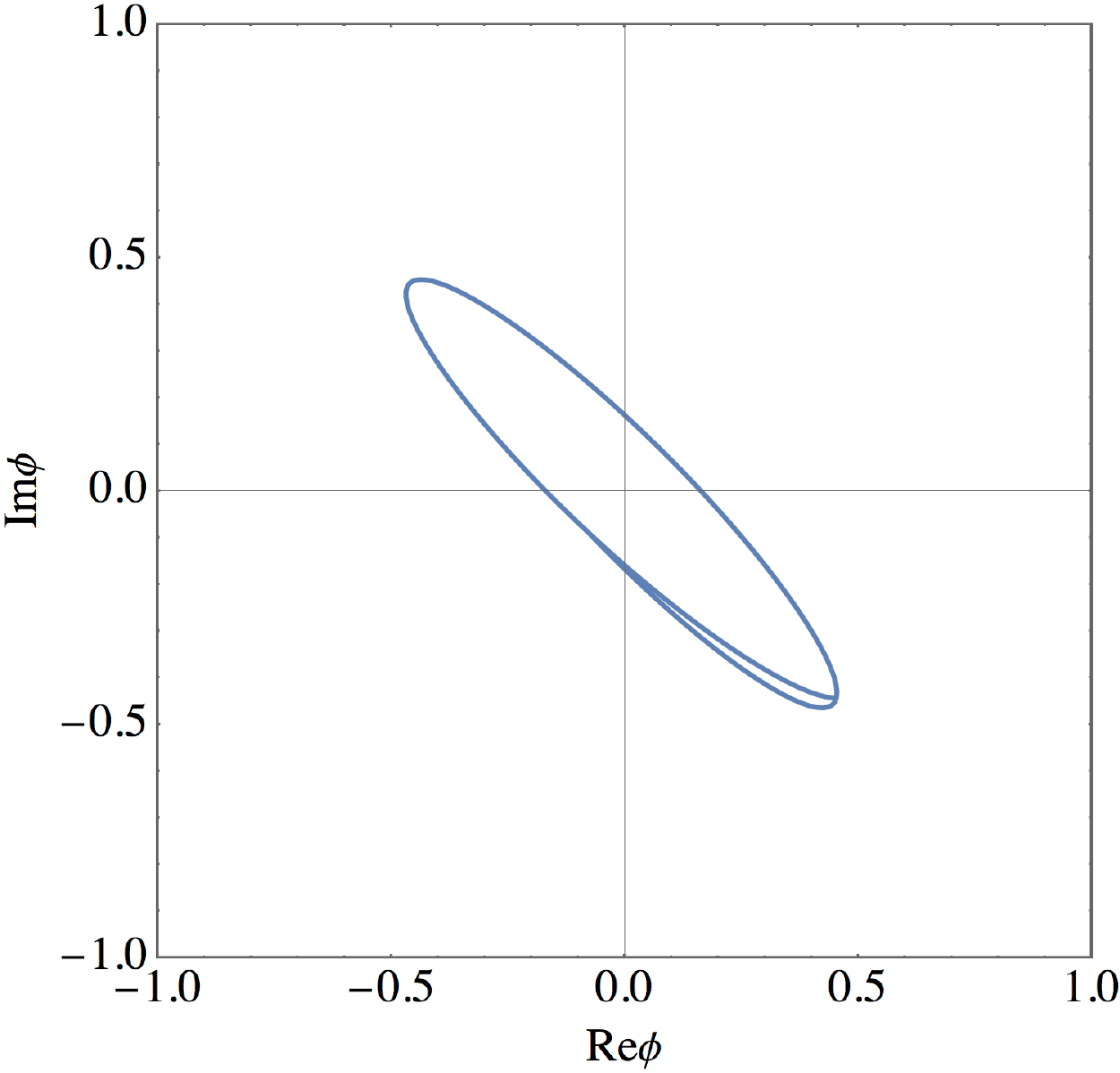}
 \end{minipage}
  \hspace{0.5cm}  
  \begin{minipage}{.45\linewidth}
  \includegraphics[width=40mm]{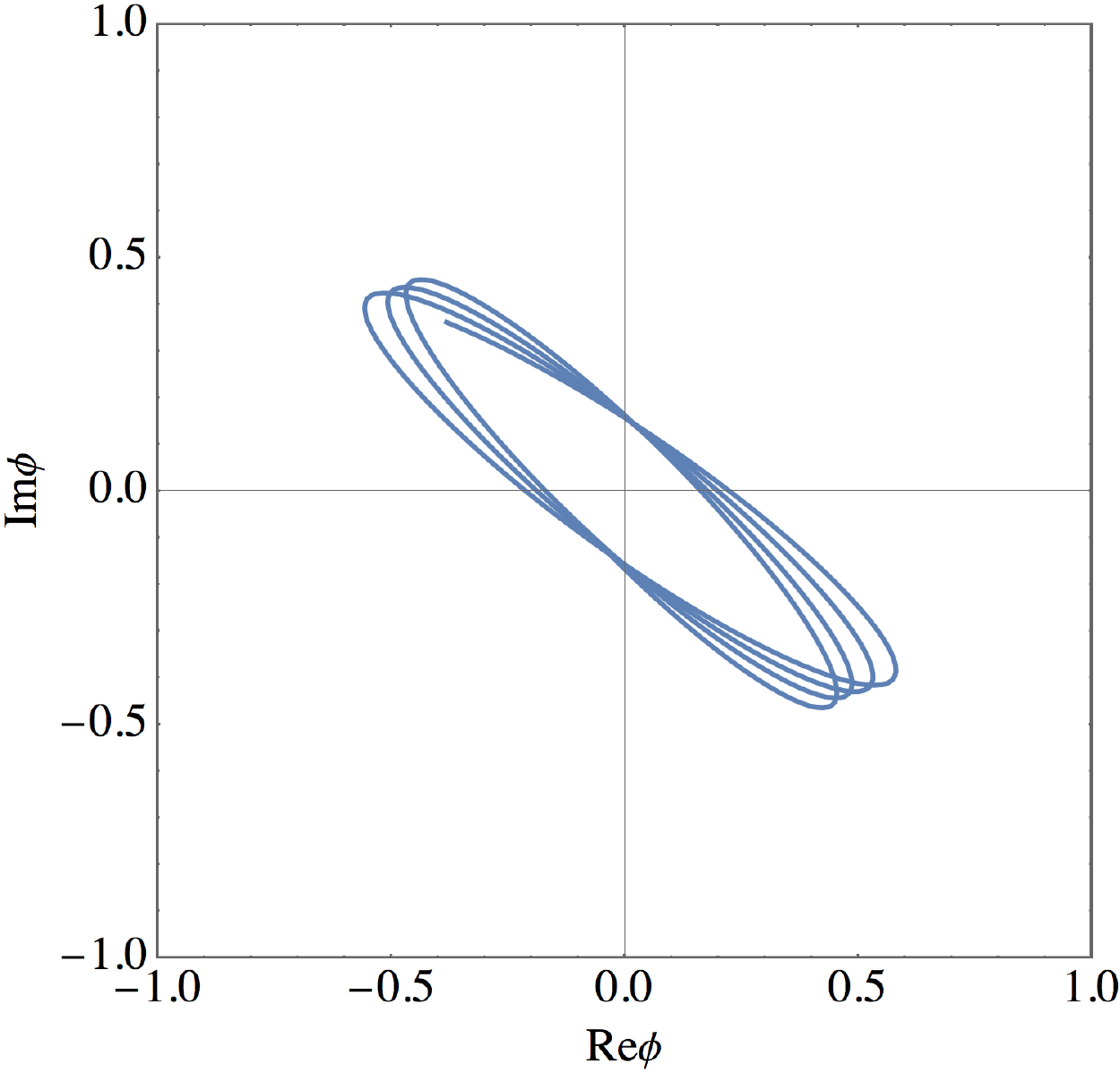}
 \end{minipage}
 \end{center}
   \caption{\sl \small  The orbit of the elliptical Q-ball produced in the lattice simulation for about one oscillation time (left) and about three oscillation times (right).}
   \label{fig:nor}
\end{figure}

As a result of our simulation, we confirmed that the non-linear lumps are formed at $t_{\rm form}\sim8\times10^2$, which are stable, that is, do not decay into the scalar waves, until the final time. We present the result in Fig.~\ref{fig:3dp}, where we set $\epsilon_{\rm ini}=0.2$. From the figure, we can see that the non-linear spherical objects are formed. We will identify these objects as the excited Q-balls by fitting the approximate semi-analytic profiles as we will discuss later.
First of all, let us confirm that the produced Q-ball has an elliptical orbit.
In Fig.~\ref{fig:nor}, we present the orbit at the center of the lump, which is plotted around the final time $t_f=5000$. We can see that it is indeed an ellipse that has a rotating axis (left figure). We also find that the orbit itself slightly rotates around the origin when we observe the orbit for a longer time (right figure), whose behavior will be explained in the next section. The normalization of the figures is the same as Eq.~(\ref{eq:norma}).

\begin{figure}[ht]
 \label{fig:spatialprofile}
      \begin{center}
   \hspace{0.5cm}  
  \begin{minipage}{.4\linewidth}
  \includegraphics[width=40mm]{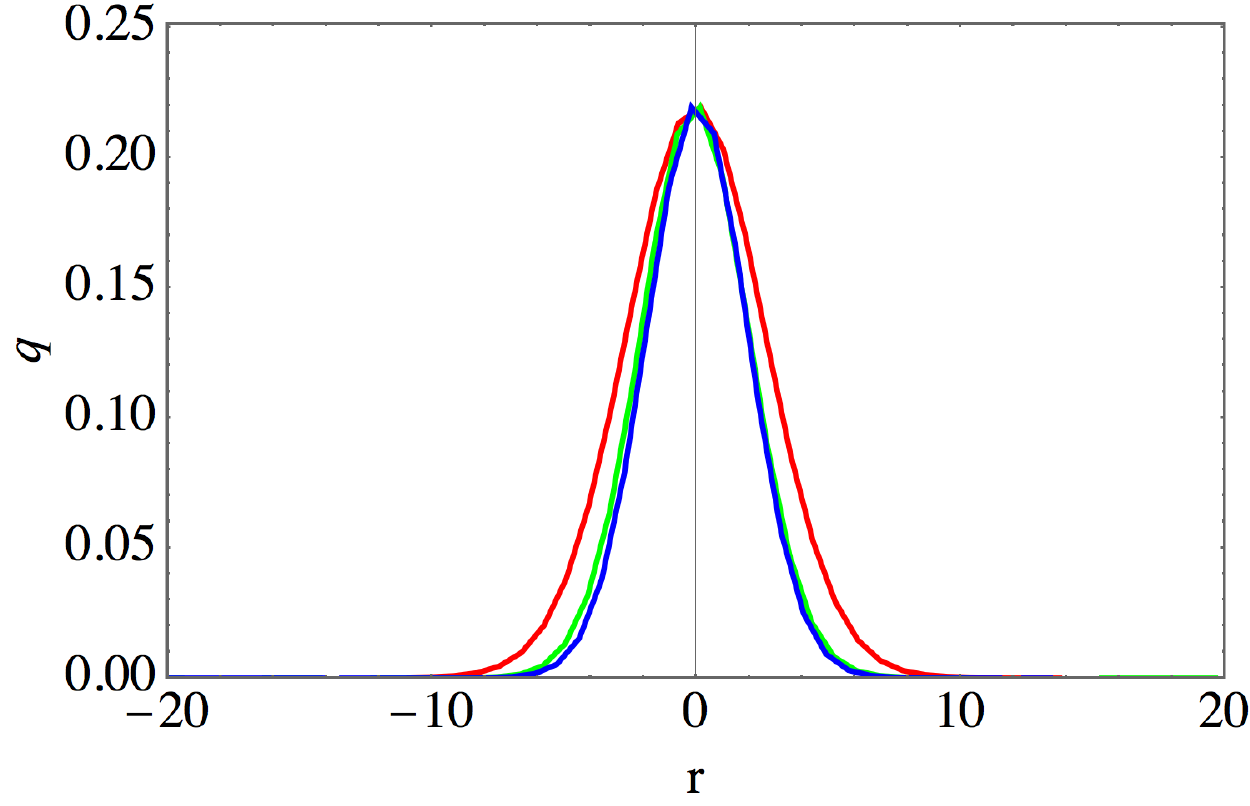}
 \end{minipage}
 \end{center}
   \caption{\sl \small  The charge density profile for the Q-ball in Fig.~\ref{fig:nor}. The labels $r$ denote the three-dimensional space coordinates rescaled to the physical distances. 
   }
   \label{fig:qx}
\end{figure}

In Fig.~\ref{fig:qx}, we also present the charge density profile of the lump with the maximal charge. The horizontal axes denote the spatial distance from the Q-ball center. The vertical axis is the charge density. The red, green, and blue curves show the charge density profile for the $x$, $y$, and $z$ coordinates. It should be noted that the charge density profile of the elliptical Q-ball is almost spherically symmetric.
Such spherical profiles of the Q-balls are also seen in Fig.~\ref{fig:3dp}. We also note that there are no nodes for the profile. Those properties suggest that the elliptical Q-ball is different from the excited states discussed in Refs.~\cite{Volkov:2002aj,Polyakov:2018zvc}.%
\footnote{One may wonder that the charge density profile slightly deviates from the spherically symmetric profile, and thus it is related to the excited state of the Q-balls. As we show in the following, the dispersion relation of elliptical Q-ball is almost determined by a single parameter, $i.e.$ the ellipticity of the orbit. Furthermore, the approximate solution for the elliptical Q-ball under the spherically symmetric spatial profile ansatz describes the elliptical Q-ball properties well.
Therefore, we conclude that the elliptical Q-ball is not the same as the excited state discussed in Refs.~\cite{Volkov:2002aj,Polyakov:2018zvc}, while the slight deviation from the spherical symmetry might be explained through the perturbative angular excitation of the Q-balls.}

\begin{figure}[t]
 \label{fig:dispersion}
       \begin{flushleft}
  \begin{minipage}{\linewidth}
  \includegraphics[width=70mm]{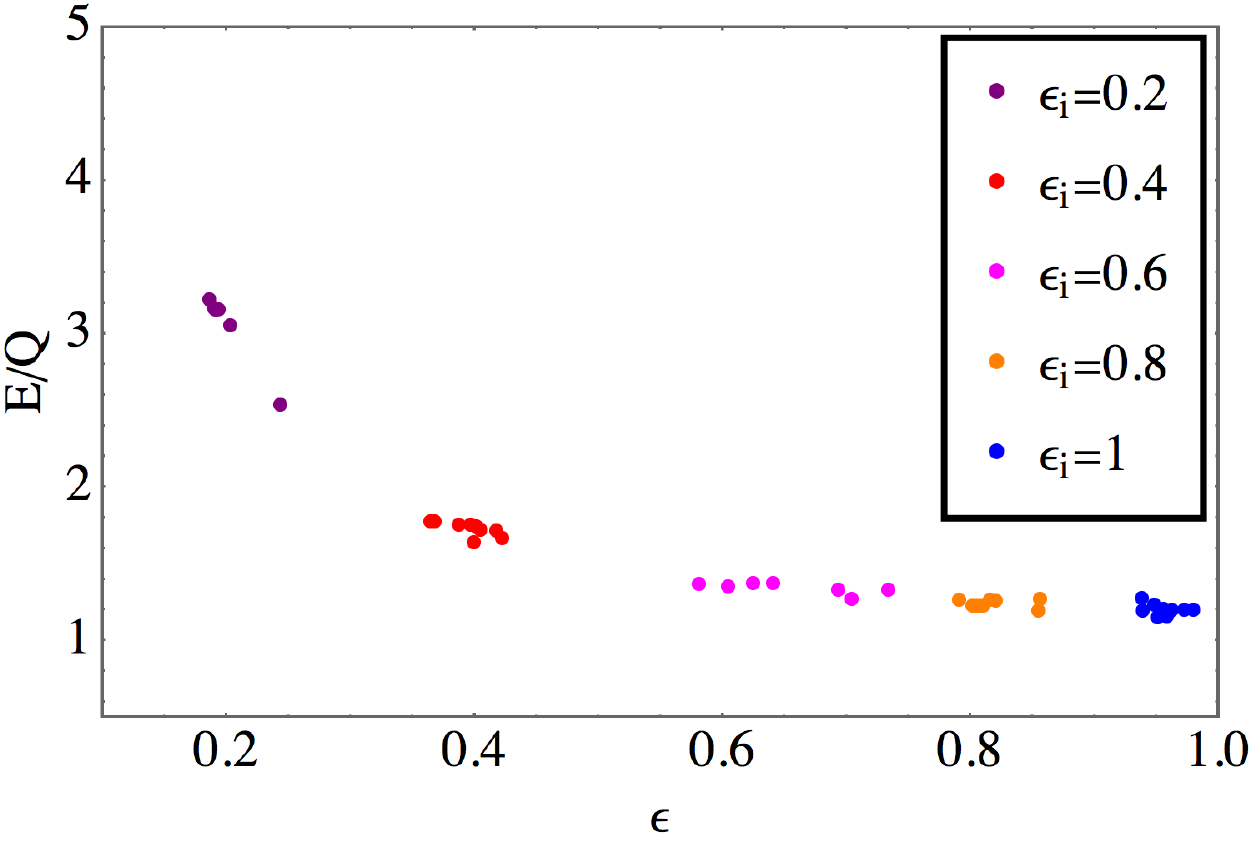}
 \end{minipage}
 \end{flushleft}
 \caption{\sl \small  The dispersion relation $E/Q$ obtained from the Q-ball formed in the lattice simulation. 
 The difference in color denotes the difference in the initial parameter $\epsilon_{\rm ini}$. }
   \label{fig:disp1}
\end{figure}

Next, let us see the dispersion relation of the elliptical Q-ball.
In Fig.~\ref{fig:disp1}, we plot the relation between $E/Q$ and $\epsilon$ of the Q-balls with the charge of $10\lesssim Q\lesssim100$, where $\epsilon$ is defined by the short axis of the ellipse divided by the long axis of the ellipse. The difference in the color of the dots denotes that in the initial condition. We vary the initial ellipticity $\epsilon_{\rm ini}$ from $0.2$ to $1$. 

We confirm that $E/Q$ of the elliptical Q-balls are larger than those of the ground state Q-balls. In other words, it is shown that the energy of the elliptical Q-balls is larger than that of the ground state Q-balls for a fixed charge. This time we have run the simulations with $N_{\rm grid}=64^3$ in order to save the time of computation. We confirmed for several samples that there is no essential difference from the case of $N_{\rm grid}=128^3$ in the plot.

\section{Approximate solution for elliptical Q-ball}
\label{sec:5}

In this section, we derive the approximate solution for the elliptical Q-ball. We consider two different approaches to obtain such a solution. Those approximations lead to the dispersion relation of the elliptical Q-balls which roughly fits the simulation results.

\subsection{Extended Gaussian Approximation}
The ground state Q-ball solution in gravity mediation is well fitted by the Gaussian approximation. The elliptical Q-ball solution is also approximated by considering the extension of the Gaussian profile as explained below.

Let us take the Gaussian ansatz,
\begin{align}
\phi(t,{\bf x})=\varphi(t) e^{-r^2/R^2},
\end{align}
where $\varphi(t)$ is a complex function of $t$. 
We decompose $\varphi(t)$ into the radial direction $\Phi(t)$ and the phase direction $\theta(t)$,
\begin{align}
\label{eq:1an}
\varphi(t)=\frac{1}{\sqrt{2}}\Phi(t)e^{i\theta(t)}.
\end{align}
Here, we do not specify the time-dependence of $\Phi(t)$ and $\theta(t)$. We note that the time dependence of  $\varphi(t)$ in the ordinary Gaussian ansatz for the ground state Q-ball solution is just
\begin{align}
\varphi(t)\sim e^{i\omega t}
\end{align}
with constant $\omega$. 

From the ansatz in Eq.\,\eqref{eq:1an}, the energy of the system is rewritten as
\begin{align}
E=\int d^3x\left[\frac{1}{2}\dot\Phi^2+\frac{1}{2}\dot\theta^2\Phi^2+\frac{1}{2}\left(\frac{2r}{R^2}\right)^2\Phi^2\right.\nonumber\\
\left.+\frac{1}{2}\Phi^2\left[1+K\ln\left({\Phi^2e^{-2r^2/R^2}}\right)\right]\right]e^{-2r^2/R^2}.
\end{align}
Performing the integration for the spatial coordinate, we obtain the energy formula,
\begin{align}
E=\frac{1}{2\sqrt{2}}\pi^{3/2}R^3\left[\frac{1}{2}\dot\Phi^2+\frac{1}{2}\dot\theta^2\Phi^2+\frac{1}{2R^2}\Phi^2\right.\nonumber\\
\left.+\frac{1}{2}\Phi^2\left[1+K\ln\left({\Phi^2e^{-3/2}}\right)\right]\right].
\end{align}

From the $U(1)$ charge conservation, the following relation is also obtained,
\begin{align}
\label{eq:q2}
\int d^3x\dot\theta\Phi^2e^{-2r^2/R^2}=\frac{1}{2\sqrt{2}}\pi^{3/2}R^3\dot\theta\Phi^2=Q~({\rm constant}).
\end{align}

Using the above relation in Eq.~(\ref{eq:q2}), the energy of the system is represented only by $\Phi(t)$,
\begin{align}
\tilde{E}=\frac{1}{2}\dot\Phi^2+U(\Phi),
\end{align}
where $U(\Phi)$ is an effective potential,
\begin{align}
\label{eq:effectivepotential}
U(\Phi)=\frac{1}{2}\frac{\tilde{Q}^2}{\Phi^2}+\frac{1}{2R^2}\Phi^2+\frac{1}{2}\Phi^2\left[1+K\ln\left({\Phi^2e^{-3/2}}\right)\right],
\end{align}
and we define $\tilde{E}=2\sqrt{2}E/\pi^{3/2}R^3$ and $\tilde{Q}=2\sqrt{2}Q/\pi^{3/2}R^3$.

\begin{figure}[t]
   \centering
   \includegraphics[width=70mm]{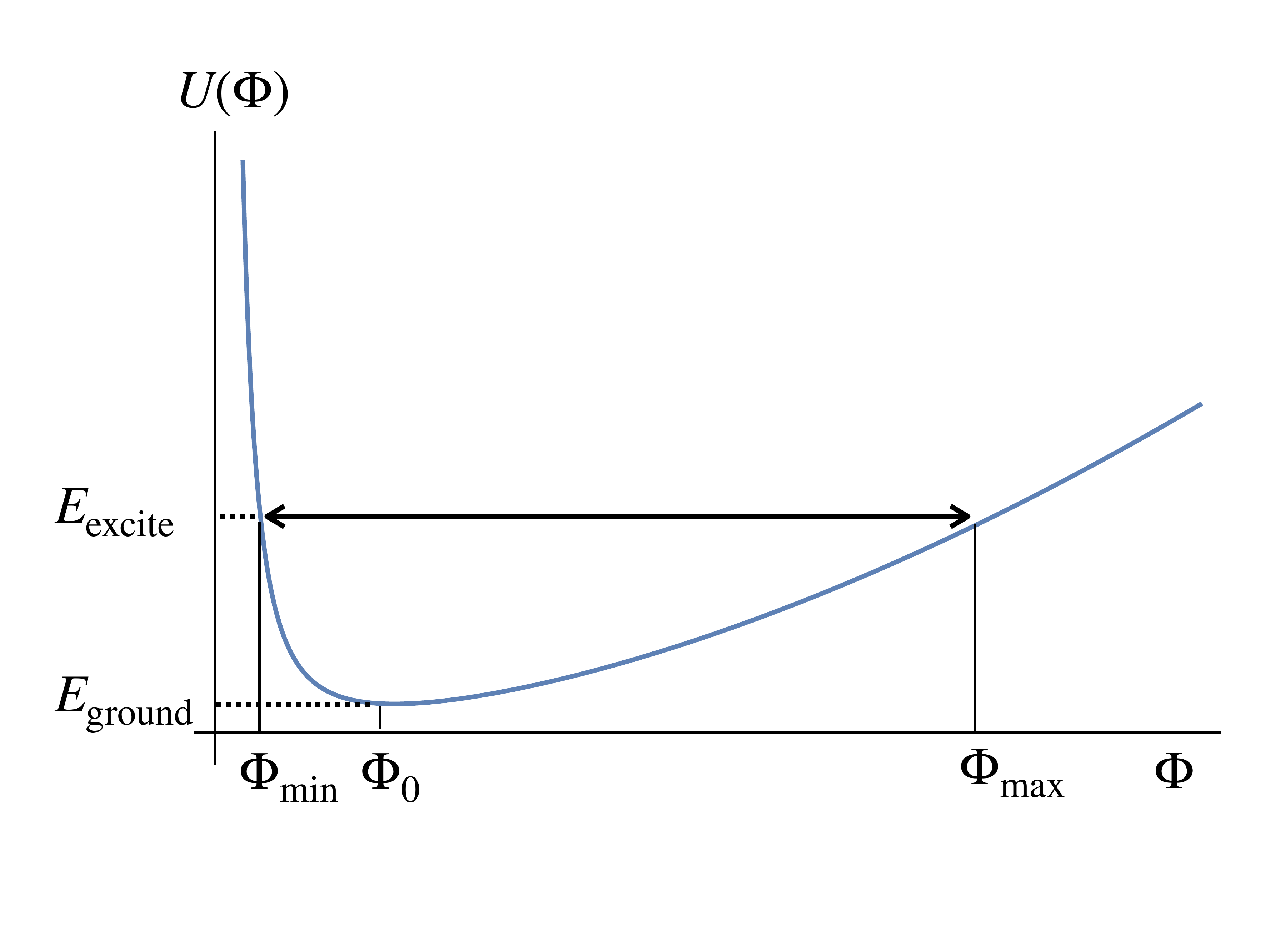}
   \caption{\sl \small  The effective potential for $\Phi$.
   } 
   \label{fig:ep}
\end{figure}

This system is analogous to that of Keplerian motion of the planet in the gravitational potential.
The lowest energy state described in this system is apparently~(Fig.~\ref{fig:ep}),
\begin{align}
({\rm Ground\ state})~~\dot\Phi=0,~\Phi=\Phi_0\simeq \tilde{Q}^{1/2},
\end{align}
where $\Phi_0$ is a minimum of the effective potential and we used $R^{-2}\sim|K|\ll 1$.
In this case, $\dot\theta=Q/\Phi_0^2=$ constant and $\varphi$ describes the circular orbit in the complex plane.
This solution is nothing but the ground-state Q-ball solution.
Then, the energy per unit charge is given by
\begin{align}
E_{\rm ground}/Q=U(\Phi_0)/Q\simeq 1,
\end{align}
which is consistent with the result of a conventional formulation of the Q-ball.

\begin{figure}[t]
   \centering
   \includegraphics[width=60mm]{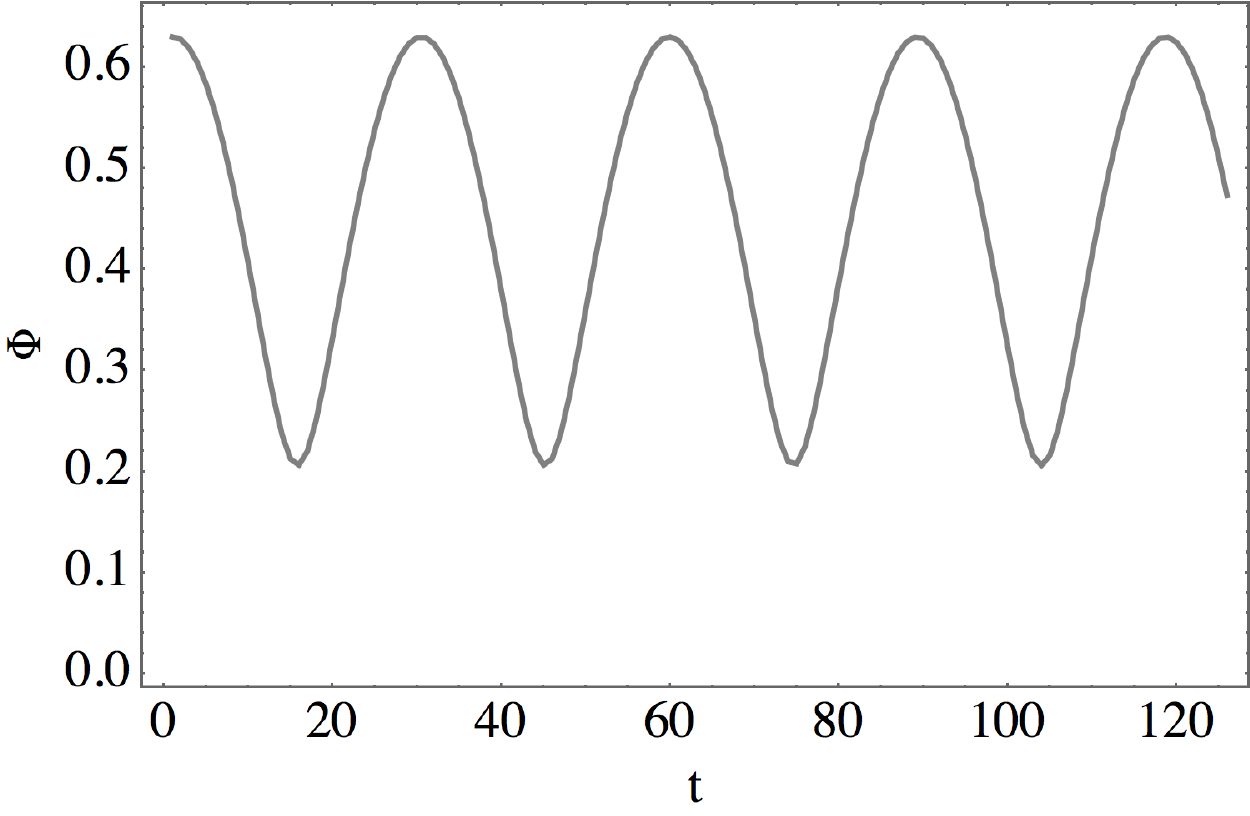}
   \caption{\sl \small  The oscillation of $\Phi$ for $Q=50,~K=-0.1,~R=\sqrt{2/|K|}$.
   } 
   \label{fig:po}
\end{figure}

To obtain the excited state solution, let us consider the case $\dot\Phi\neq0$ in the system. Because $\Phi$ obtains the kinetic energy, $\Phi$ oscillates around the potential minimum $\Phi_0$ (Fig.~\ref{fig:po}).

In such a condition, the radial direction of the complex field $\phi$ oscillates between $\Phi_{\rm min}$ and $\Phi_{\rm max}$, which means that the orbit of $\phi$ is elliptical.
Due to the non-vanishing kinetic energy of $\Phi$, the energy of this Q-ball is larger than circular  (ground state) Q-ball:
\begin{align}
E_{\rm excite}/Q=U(\Phi_{\rm min})/Q=U(\Phi_{\rm max})/Q\gtrsim 1.
\end{align}

Since $\Phi_{\rm min}$ and $\Phi_{\rm max}$ are related to the ellipticity parameter as $\epsilon=\Phi_{\rm min}/\Phi_{\rm max}$, the ratio $E/Q$ is calculated as a function of $\epsilon$.
Neglecting the contribution of the order $|K|$, we obtain
\begin{align}
\tilde E_{\rm excite}&\simeq\frac{1}{2}\frac{\tilde{Q}^2}{\Phi^2}+\frac{1}{2}\Phi^2 \nonumber\\
& \Leftrightarrow\  \Phi^2=
\begin{cases}
\Phi_{\rm max}^2=\tilde E_{\rm excite}^2+\sqrt{\tilde E_{\rm excite}^2-\tilde{Q}^2}\\
\Phi_{\rm min}^2=\tilde E_{\rm excite}^2-\sqrt{\tilde E_{\rm excite}^2-\tilde{Q}^2}.
\end{cases}
\end{align}
Therefore,
\begin{align}\label{drom}
\epsilon^2&=\frac{\tilde E_{\rm excite}^2-\sqrt{\tilde E_{\rm excite}^2-\tilde{Q}^2}}{\tilde E_{\rm excite}^2+\sqrt{\tilde E_{\rm excite}^2-\tilde{Q}^2}} \nonumber\\
\Rightarrow\ &  E_{\rm excite}/Q\simeq\frac{1+\epsilon^2}{2\epsilon}.
\end{align}
While we have neglected the $K$ dependence in the above formulas for simplicity, for the actual computation using our approximate analysis we are solving the field equation including the $K$ dependence. However, it will turn out that the $K$ dependence in the dispersion relation, for instance, is negligible for $K\lesssim 0.1$, so that the above formula is actually allowed in practice. We will discuss on this point at the end of this subsection.

 \begin{figure}[t]
   \centering
   \includegraphics[width=60mm]{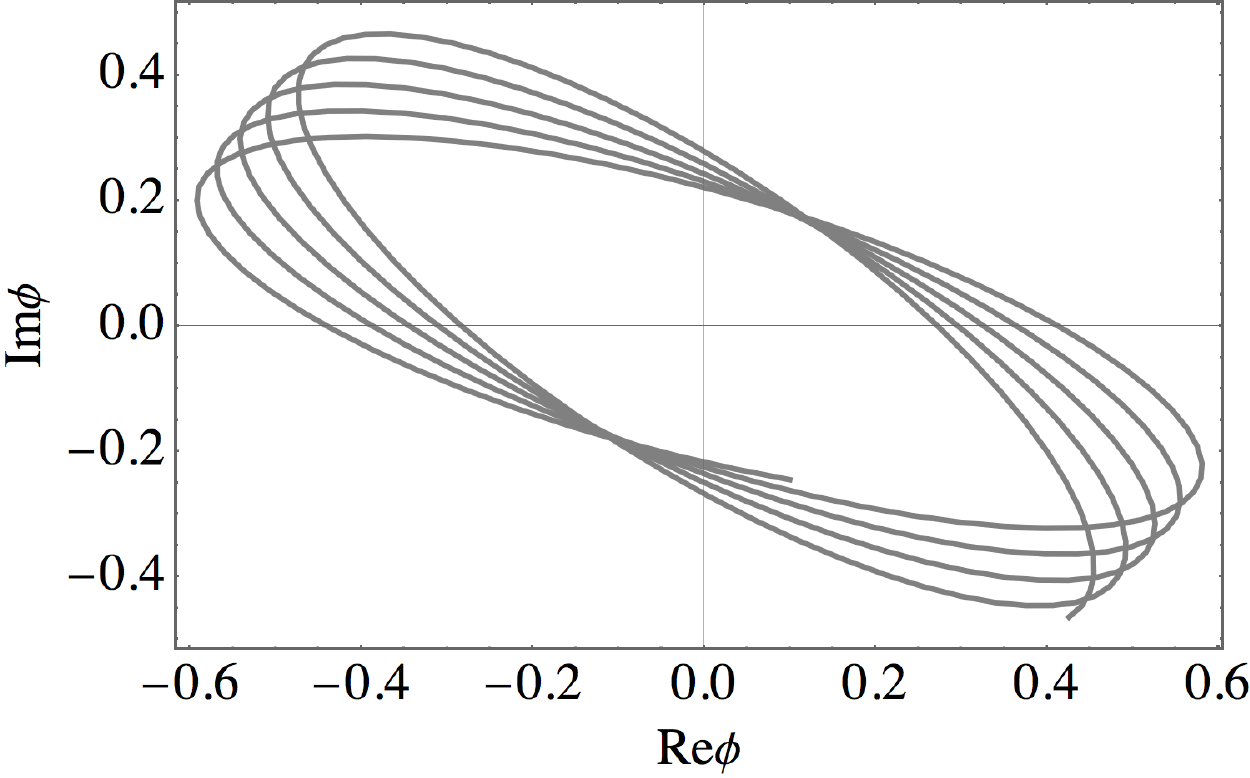}
   \caption{\sl \small  The orbit of $\varphi$ for the excited Q-ball for $Q=50,~R=\sqrt{2/|K|}$.
   } 
   \label{fig:or}
\end{figure}

The dispersion relation in Eq.~(\ref{drom}) is an extension from that of circular Q-ball. 
While elliptical Q-ball solution coincides with the circular Q-ball solution for $\epsilon=1$, it carries larger energy per unit charge than the circular one for $\epsilon\neq1$.
Therefore, in this sense, we can regard the elliptical Q-ball as a kind of ``excited state" of the Q-ball.

Furthermore, we also show the orbit of $\phi$ in the complex plane in Fig.~\ref{fig:or}. There, we take $Q=50,~R=\sqrt{2/|K|}$. We can see that the axis of the ellipse is rotating in time. This behavior seems to agree with the simulation results in Fig.~\ref{fig:nor}.

\subsubsection*{Comparing with Simulation Results}
\begin{figure}[t]
      \begin{flushleft}
  \begin{minipage}{.4\linewidth}
   \includegraphics[width=40mm]{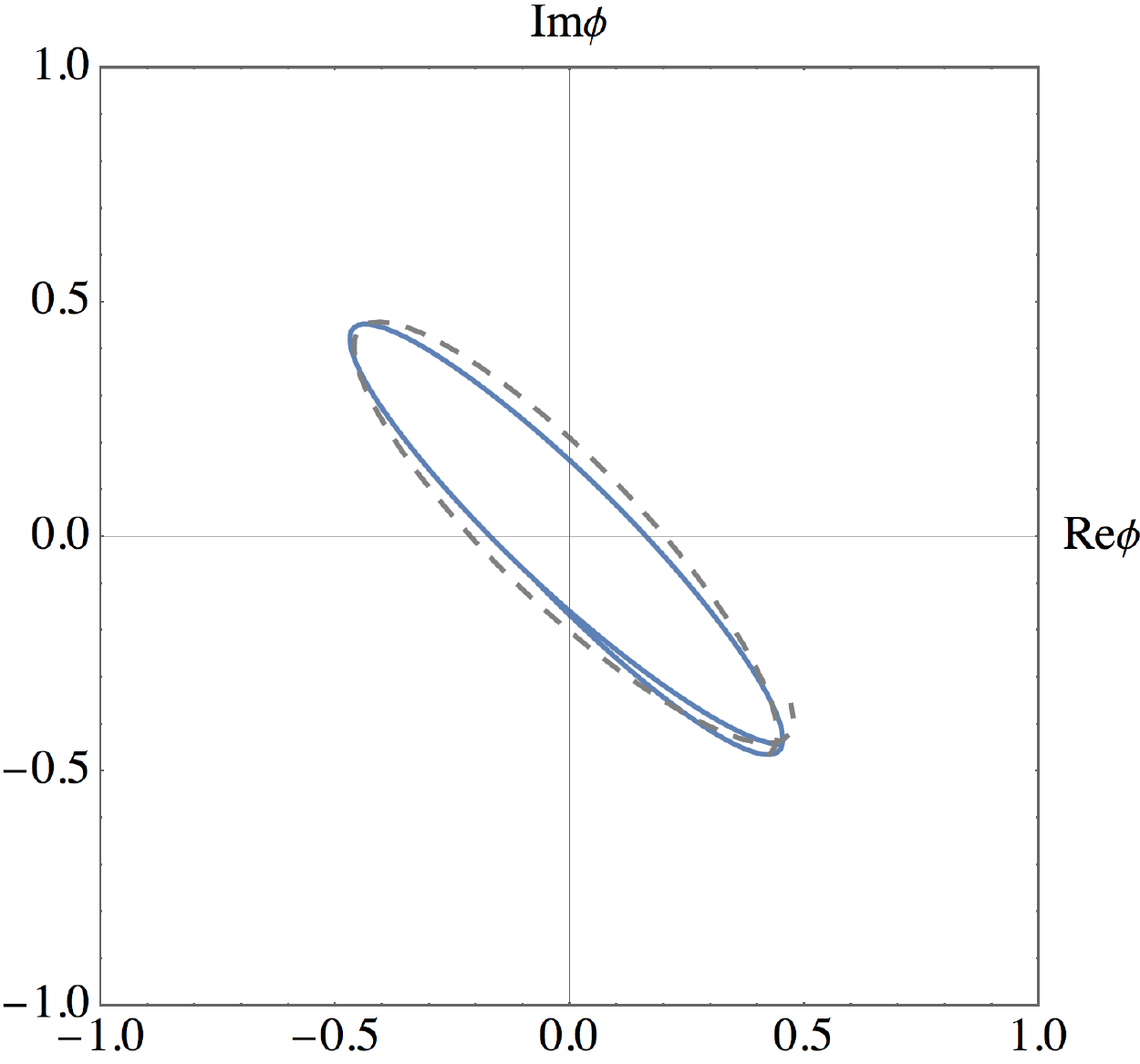}
 \end{minipage}
  \hspace{0.5cm}  
  \begin{minipage}{.4\linewidth}
    \includegraphics[width=45mm]{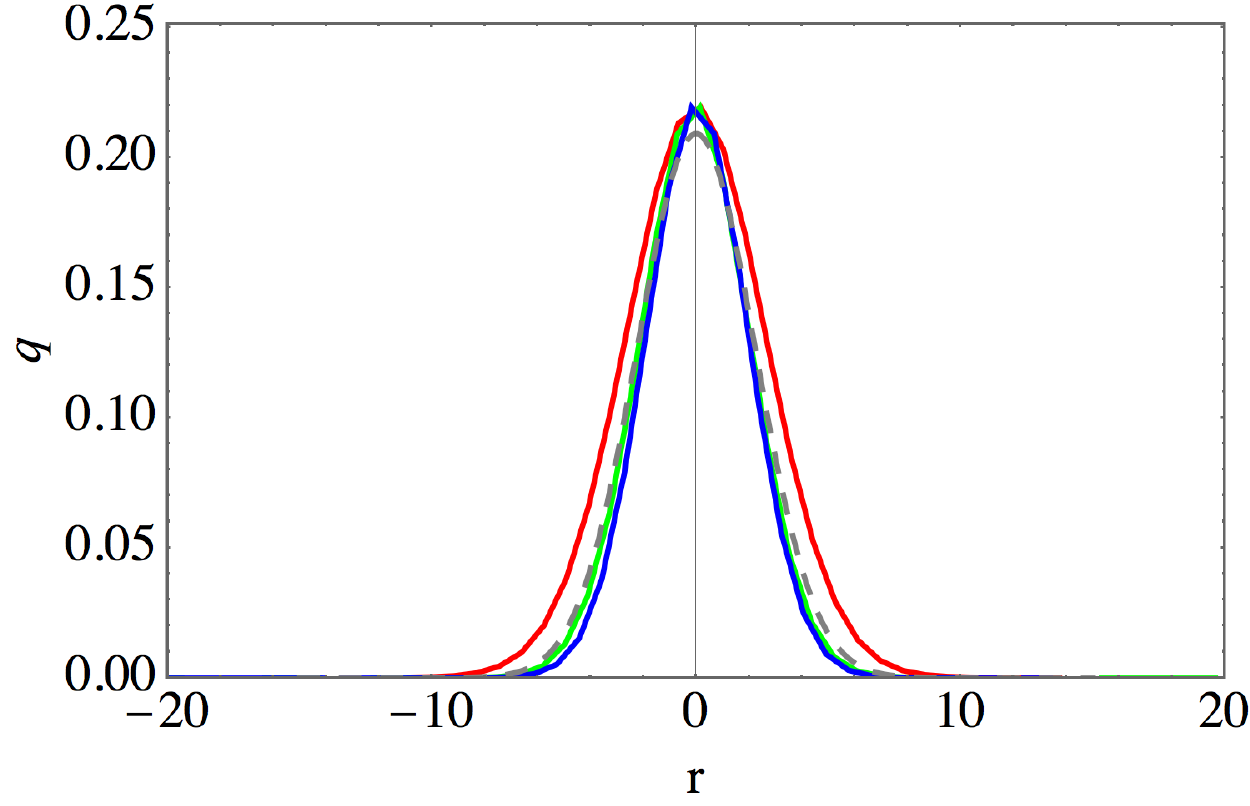}
 \end{minipage}
   \end{flushleft}
   \caption{\sl \small  Fitting in extended Gaussian approximation. The gray dashed line is the orbit and the charge density profile from the extended Gaussian approximation in the left and the right figures, respectively. The solid lines are the same as those in Fig.~\ref{fig:nor}~(left) and Fig.~\ref{fig:qx}. Here, we take $Q=36.5,~R=\sqrt{2/|K|}$.}
   \label{fig:qx3}
\end{figure}

To confirm the availability of the extended Gaussian approximation, let us compare the approximation with the simulation results. In Fig.~\ref{fig:qx3}, we show the fitted lines for the orbit and the charge density.
The gray dashed lines are the results by the extended Gaussian approximation with $Q=36.5,~R=\sqrt{2/|K|}$.%
\footnote{For the fitting of the extended Gaussian method, we just use the radius $R=\sqrt{2/|K|}$ in Eq.~(\ref{eq:R}). }
The other color lines are the same as the orbit and the charge density in Fig.~\ref{fig:nor} and Fig.~\ref{fig:qx}. We confirm the rough agreement between the approximation and the simulation results.
More quantitatively, we can obtain the parameters $Q,~\Phi_{\rm max},~\epsilon$ of the elliptical Q-ball from the simulations.
On the other hand, we solve the field equation by inputting the parameters $Q,~R,~\Phi_{\rm max}$ and obtain $\epsilon$ in our approximate analysis.
Then, we can compare $\epsilon$ from the simulation with $\epsilon$ from the approximate analysis, in order to examine the validity of the latter.
We present the results for several samples in Tab.~\ref{tab:1}. The table shows that our approximate theory is valid for these samples with about $10$\% precision level. We especially note that the slight deviation from the spherical symmetry does not seem to be that important in the description of the properties of the elliptical Q-ball, since the approximation $R=1/\sqrt{2|K|}$ is roughly valid. For the better approximation, we may also need to take this deviation in the profile into account. We will leave the detailed analysis to the future work.

 \begin{table}[htb]
 \caption{\sl Comparing the ellipticity from the simulation $\epsilon$ with that from the approximation $\epsilon_{\rm appr}$. We use $R=\sqrt{2/|K|}$ for the approximation.}
 \begin{center}
  \begin{tabular}{|c|c|c|c|c|c|c|c|c|c|c|c|}
\hline 
$Q$ & $14$ & $34$ & $25$ & $73$ & $44$ & $110$ & $56$ & $150$ \\\hline
$\Phi_{\rm max}$ & $0.57$ & $0.99$ & $0.52$ & $0.93$ & $0.62$ & $1.0$ & $0.58$ & $1.0$ \\\hline
$\epsilon$ & $0.20$ & $0.20$ & $0.41$ & $0.38$ & $0.61$ & $0.58$ & $0.83$ & $0.80$ \\\hline 
$\epsilon_{\rm appr}$ & $0.22$ & $0.18$ & $0.47$  & $0.44$ & $0.58$ & $0.58$ & $0.86$  & $0.79$  \\\hline 
  \end{tabular}
  \end{center}
  \label{tab:1}
\end{table}

\begin{figure}[t] 
 \begin{flushleft}
  \begin{minipage}{\linewidth}
  \includegraphics[width=70mm]{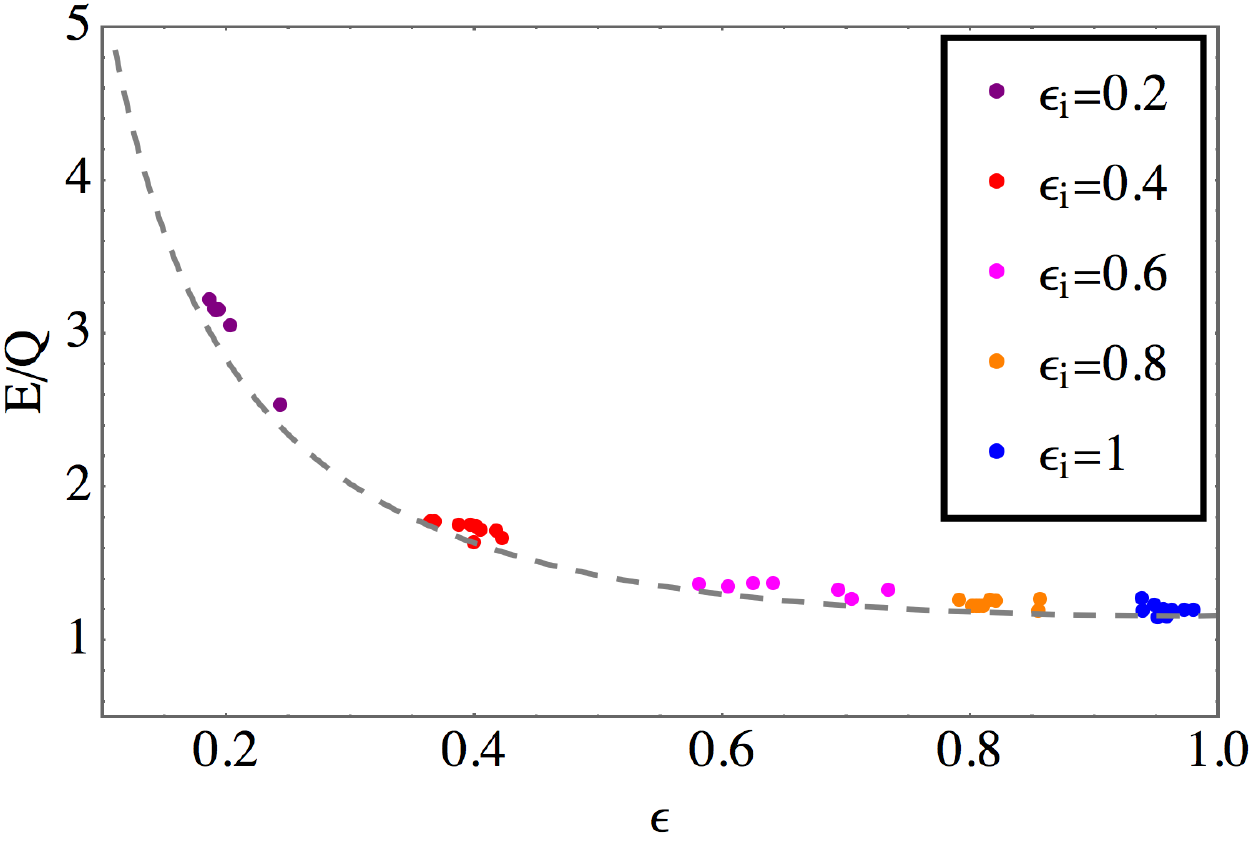}
 \end{minipage}
 \end{flushleft}
 \caption{\sl \small  The dispersion relations obtained from the lattice simulation and the extended Gaussian approximation. The gray dashed line is $E/Q$ for $Q=50$. The data from the simulation is the same as that in Fig.~\ref{fig:disp1}. Here, we take $R=\sqrt{2/|K|}$.
   }
   \label{fig:disp22}
\end{figure}

\begin{figure}[t] 
 \begin{flushleft}
  \begin{minipage}{\linewidth}
  \includegraphics[width=70mm]{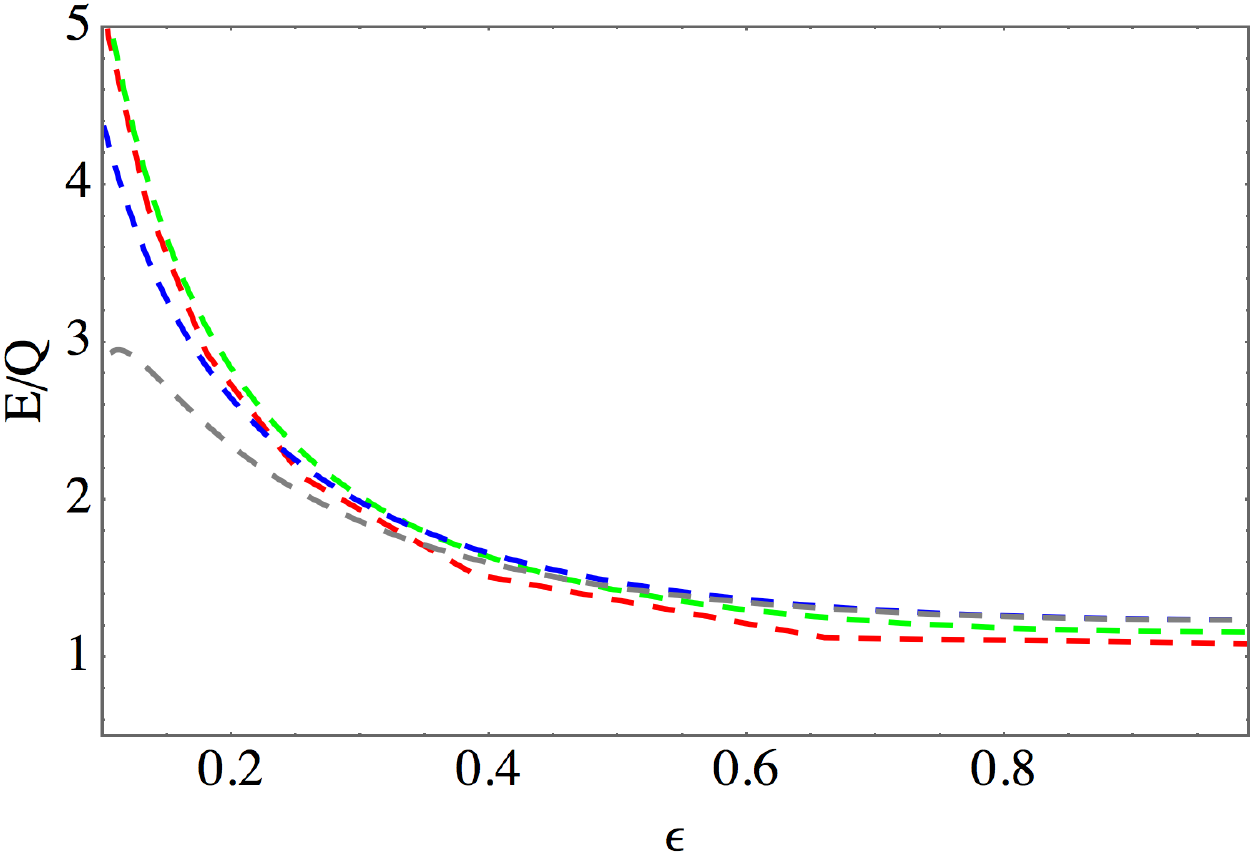}
 \end{minipage}
 \end{flushleft}
 \caption{\sl \small  The $K$ dependence of the dispersion relations. The red, green, blue, gray dashed lines are the dispersion relation for $K=-0.01,-0.1,-0.3,-0.4$. Here, we take $Q=50,~R=\sqrt{2/|K|}$.}
   \label{fig:eqkd}
\end{figure}

Using the approximation, the dispersion relation $E/Q$ for given $\epsilon$ is also calculated in Fig.~\ref{fig:disp22}.
The gray dashed line is the result for $Q=50$ (and $R=\sqrt{2/|K|}$), which fits well with the simulation results. Note that we are solving the field equation including the $K$ dependence to draw the gray dashed line, as we mentioned earlier. However, the $K$ dependence is almost negligible for $K\lesssim 0.1$, so that we can use the simple formula Eq.~(\ref{drom}) in practice: See the $K$ dependence of the dispersion relation in Fig.~\ref{fig:eqkd}.%
\footnote{We also confirmed that the $K$ dependence is almost negligible for several different $Q$ values.}
 
\subsection{Complete Ellipse Approximation}
As another way, let us consider the time-averaged equation of motion like the I-ball/oscillon solution. As shown in Fig.~\ref{fig:or}, the orbit of the elliptical Q-ball is almost closed during one oscillation. Focusing on such adiabatic motion of the elliptical Q-ball, we take the complete elliptical orbit ansatz%
\footnote{The elliptical orbit is not completely closed and slightly rotating as discussed in the previous section.},
\begin{align}\label{eor}
\phi=\frac{1}{\sqrt2}\Phi({\bf x})P(\omega t),~~~P(\omega t)=\sqrt{\frac{1}{\epsilon}}(\cos\omega t+i\epsilon\sin\omega t),
\end{align}
where $\epsilon$ is the ellipticity parameter in the defined range of $0<\epsilon \leq 1$. For $\epsilon=1$, the ansatz is nothing but the one for the ground state Q-ball in Eq.~(\ref{sepa}).

Under this ansatz, the total charge $Q$ is given by
\begin{align}
Q=\int d^3x\omega\Phi({\bf x})^2.
\end{align}
Apparently, the charge is time-invariant under the above ansatz.

The time-averaged equation of motion for $\Phi$~\cite{Bogolyubsky:1976yu,Gleiser:1993pt,Kasuya:2002zs} is
\begin{align}
\label{eq:eom3}
\frac{d^2\Phi}{dr^2}+\frac{2}{r}\frac{d\Phi}{dr}+\omega^2\Phi-\frac{\partial \bar{ V}(\Phi)}{\partial \Phi}=0,
\end{align}
where $\bar{ V}(\Phi)$ is the time-averaged potential,
\begin{align}
\bar{ V}(\Phi)=\frac{1}{2}\Phi^2\left(1+c(\epsilon)+K\ln\Phi^2\right).
\end{align}
Here,
\begin{align}
c(\epsilon)&=\frac{2\epsilon}{1+\epsilon^2}\frac{K}{2\pi/\omega}\int_0^{2\pi/\omega}|P(\omega t)|^2\ln\left(\frac{1}{2}|P(\omega t)|^2\right)\nonumber\\
&\simeq K\ln\left(\frac{1+\epsilon^2}{2\epsilon}\right).
\end{align}

Using the above formulas, we obtain the dispersion relation,
\begin{align}\label{drel}
\overline{E}/Q\simeq\frac{1+\epsilon^2}{2\epsilon},
\end{align}
from the time-averaged energy,
\begin{align}
\overline{E}=\frac{1+\epsilon^2}{2\epsilon}\int dx^3\left[\frac{1}{2}\omega^2\Phi^2+\frac{1}{2}(\nabla\Phi)^2\right.\nonumber\\
\left.+\frac{1}{2}\Phi^2\left(1+c(\epsilon)+K\ln\Phi^2\right)\right].
\end{align}
This result is similar to the results from the extended Gaussian ansatz in Eq.\,\eqref{drom}.

\subsubsection*{Comparing with Simulation Results}

 \begin{figure}[t]
      \begin{flushleft}
  \begin{minipage}{.4\linewidth}
  \includegraphics[width=40mm]{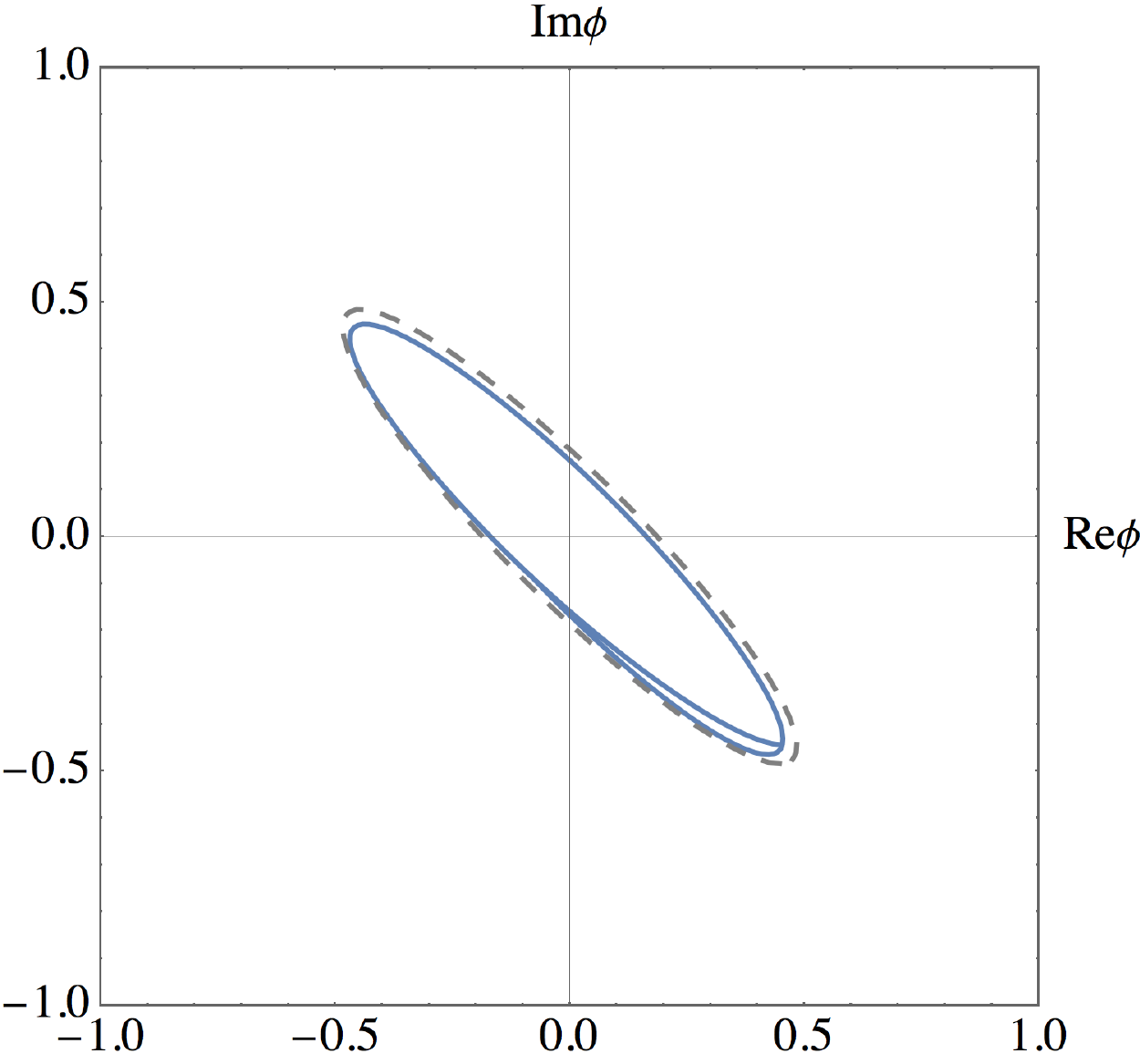}
 \end{minipage}
  \hspace{0.5cm}  
  \begin{minipage}{.4\linewidth}
  \includegraphics[width=45mm]{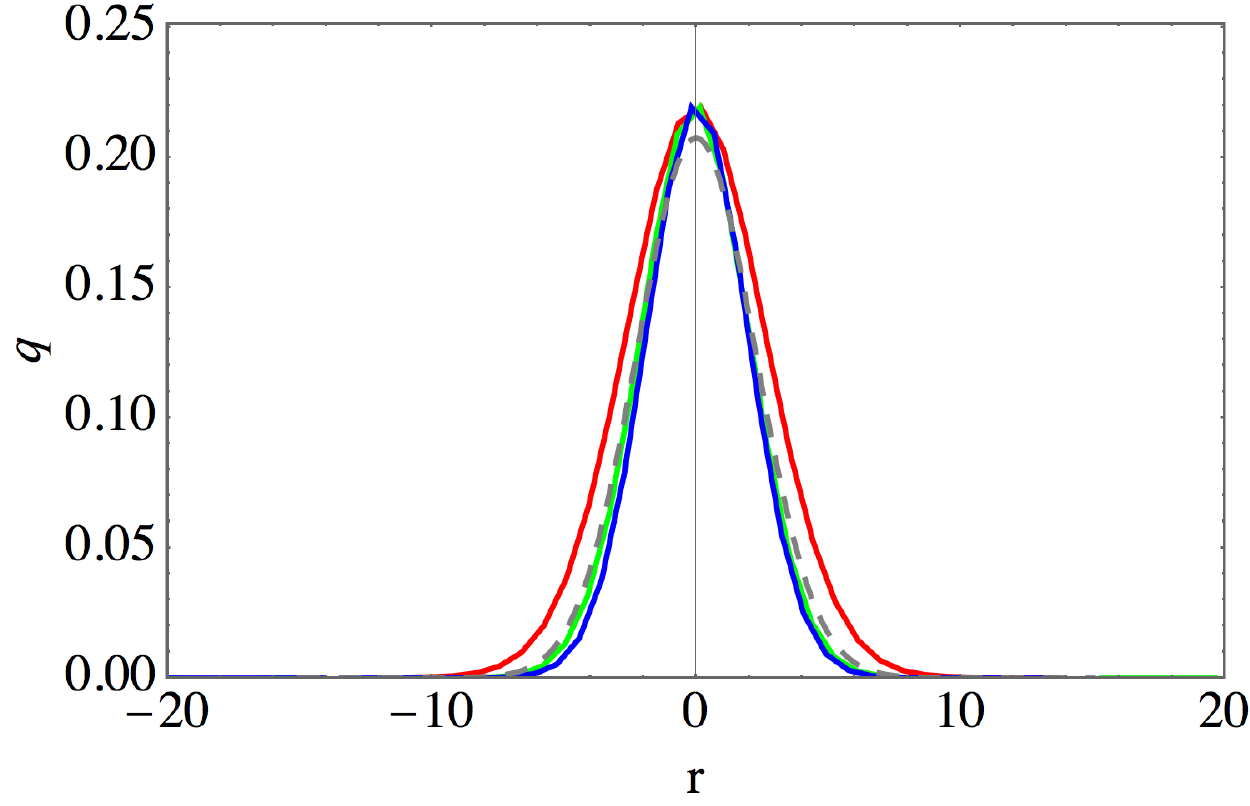}
 \end{minipage}
   \end{flushleft}
   \caption{\sl \small  Fitting in the complete ellipse approximation. The gray dashed line is the orbit and the charge density profile in the left and the right figures, respectively. The solid lines are the same as the one in Fig.~\ref{fig:nor}~(left) and Fig.~\ref{fig:qx}. Here, we take $\epsilon=0.2,~\omega=1.149$.}
   \label{fig:qx2}
\end{figure}

 \begin{figure}[t]
       \begin{flushleft}
  \begin{minipage}{\linewidth}
  \includegraphics[width=70mm]{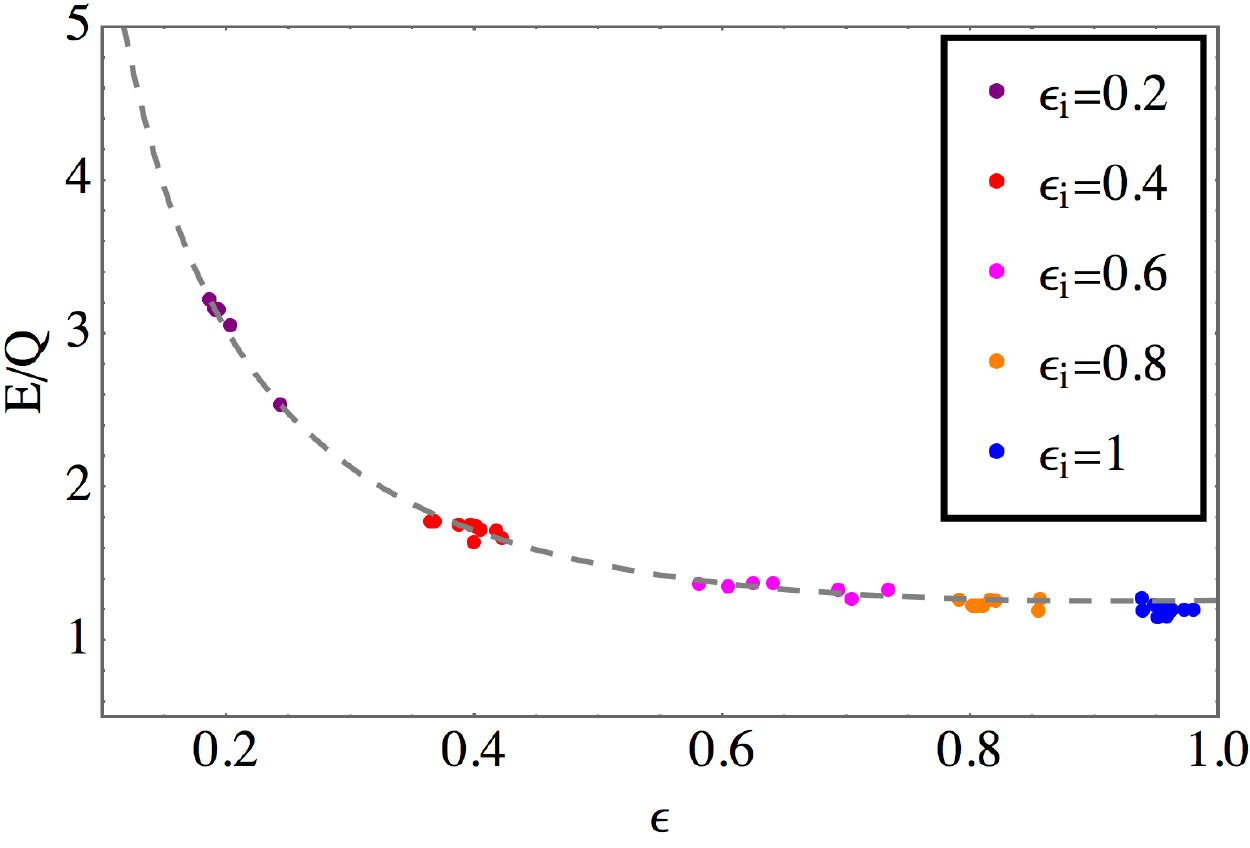}
 \end{minipage}
 \end{flushleft}
 \caption{\sl \small  The dispersion relations obtained from the lattice simulation and the complete ellipse approximation. The gray dashed line is $E/Q$ for $Q=50$. The data from the simulation is the same as the ones in Fig.~\ref{fig:disp1}. 
   }
   \label{fig:disp21}
\end{figure}

Let us first determine the variables $\omega$ and $\epsilon$ from the orbit in Fig.~\ref{fig:nor}. From the figure, the ellipticity parameter is given as $\epsilon\simeq 0.2$. The parameter $\omega$ is determined to fit the value of the field.%
\footnote{The field value of the Q-ball center is sensitive to the value of $\omega$.}
In Fig.~\ref{fig:qx2} (the left figure), we present the fitted line with $\epsilon=0.2,~\omega=1.149$ as the gray dashed line. The blue line is the same as the line in Fig.~\ref{fig:nor}. 

From the above solution, the charge density profile profile is calculated in Fig.~\ref{fig:qx2}. The gray dashed line is the result with the same parameters, $\epsilon=0.2,~\omega=1.149$. The blue line is the same as the line in Fig.~\ref{fig:qx}. Here we can also see the rough agreement between this approximate solution and the simulation result.

 \begin{table}[htb]
 \caption{\sl Comparing the ellipticity from the simulation $Q$ with that from the approximation $Q_{\rm appr}$.}
 \begin{center}
  \begin{tabular}{|c|c|c|c|c|c|c|c|c|c|c|c|}
\hline 
$\epsilon$ & $0.20$ & $0.20$ & $0.41$ & $0.38$ & $0.61$ & $0.58$ & $0.83$ & $0.80$ \\\hline 
$\Phi_{\rm max}$ & $0.57$ & $0.99$ & $0.52$ & $0.93$ & $0.62$ & $1.0$ & $0.58$ & $1.0$ \\\hline
$Q$ & $14$ & $34$ & $25$ & $73$ & $44$ & $110$ & $56$ & $150$
 \\\hline
$Q_{\rm appr}$ & $14$ & $39$ & $23$ & $67$ & $49$ & $120$ & $58$ & $167$ \\\hline
  \end{tabular}
  \end{center}
  \label{tab:2}
\end{table}

We can perform more quantitative examination of the agreement as in the previous subsection. First, we can obtain the parameters $\epsilon$, $\Phi_{\rm max}$, and $Q$ from the simulation results, and we input the parameters $\epsilon$, $\Phi_{\rm max}$\footnote{In the complete ellipse method, we determine the amplitude by determining $\omega$ and solving Eq.~(\ref{eq:eom3}) for a fixed $\epsilon$.} and derive $Q$ using the complete ellipse approximation, which we compare with the charge $Q$ from the simulation. The results for several samples are shown in Tab.~\ref{tab:2}. The table shows that the estimation of $Q$ using the complete ellipse method is valid for these samples at least with about $10\%$ precision level.

The dispersion relation $E/Q$ for $\epsilon$ is compared with the simulation result in Fig.~\ref{fig:disp21}. There, the gray dashed line is the result for $Q=50$, which is consistent with the simulations as well. Therefore, we find that this approximation is also useful for the discussion on the properties of the elliptical Q-ball.

\subsection{Stability of Elliptical Q-ball}
Before closing this section, let us briefly discuss the stability of the elliptical Q-balls. 
We know that the circular Q-ball $\epsilon=1$ is absolutely stable because it is the most stable state which the complex scalar can take.
In the case of the elliptical Q-balls, however, they might move to the more stable state, that is, less elliptical Q-balls. In Ref.~\cite{Hiramatsu:2010dx}, the fission of the elliptical Q-ball to the multiple Q-balls is observed, which are less elliptical than the original one.
Therefore, the elliptical Q-balls are considered as the ``transients", which appear during the formation of the circular Q-balls. The mechanism of how the elliptical Q-ball makes such transition is an open question. Here, we discuss that the growing perturbation around the elliptical Q-ball may be the seed for the transition to more stable Q-balls.

Since we now know that the simple approximations roughly describe the elliptical Q-ball, here we use the extended Gaussian approximation, and consider a little perturbation $\delta\Phi$ around the elliptical Q-ball solution:
\begin{align}
\Phi({\bf x},t)=\overline{\Phi}(t)+\delta\Phi({\bf x},t),
\end{align}
where $\overline{\Phi}$ is the background elliptical Q-ball solution which satisfies the equation of motion.
Then, the equation of motion for the perturbation $\delta\Phi$ is 
\begin{align}
\delta\ddot{\Phi}+U''(\overline\Phi)\delta\Phi=0,
\end{align}
whose solution will give how the perturbation grows.

 \begin{figure}[t]
       \begin{flushleft}
  \begin{minipage}{\linewidth}
  \includegraphics[width=70mm]{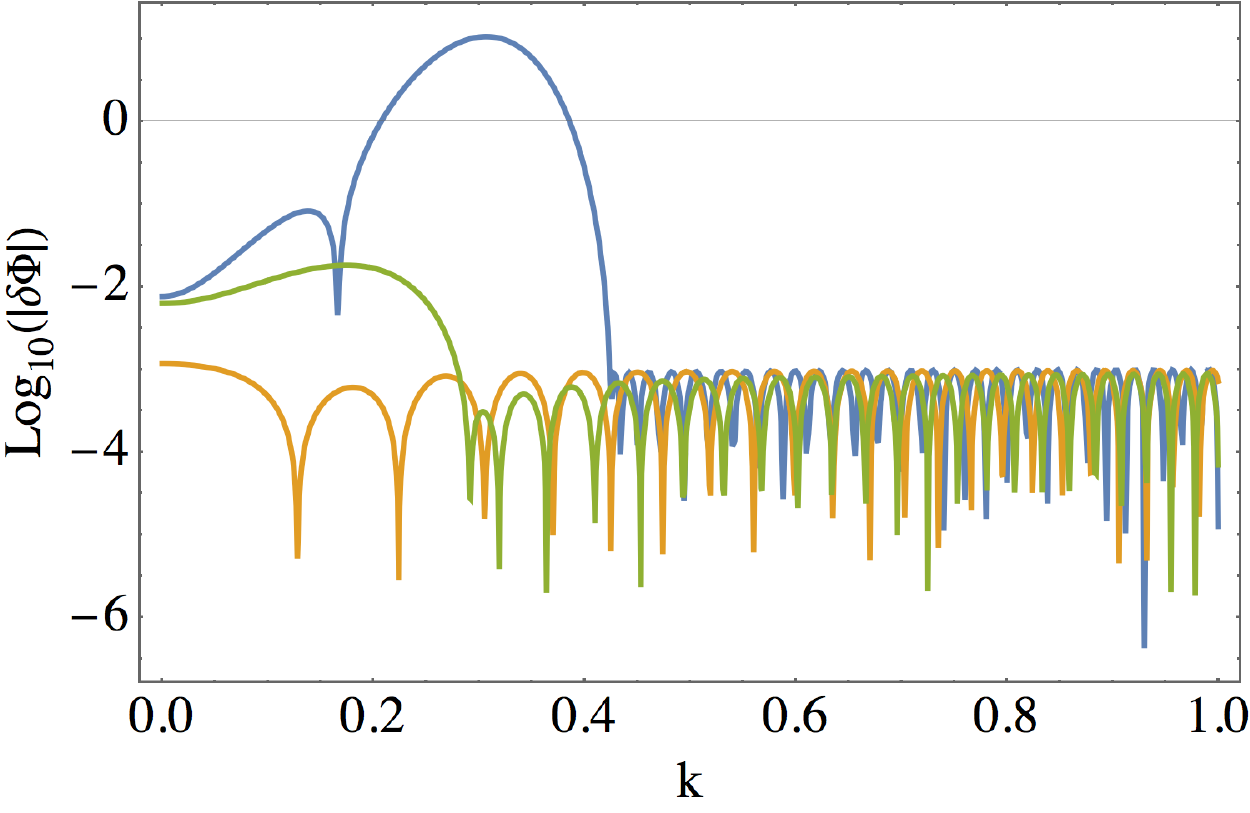}
 \end{minipage}
 \end{flushleft}
\caption{\sl \small The growth of $\delta\Phi$. The horizontal axis denotes the frequency $k$. The blue, orange, green lines correspond to the background elliptical Q-ball with $\epsilon=0.2,~0.5,~0.8$, respectively. Here, we take $Q=50,~R=\sqrt{2/|K|}$ for a demonstration.}
    \label{fig:inst}
\end{figure}

In Fig.~\ref{fig:inst}, we present the growth of the perturbation for several values of elliptical parameter $\epsilon$. Here, we take $Q=50,~R=\sqrt{2/|K|}$ for demonstration. The green, orange, blue lines denote the fluctuation $|\delta\Phi|$ for $\epsilon=0.2,~0.5,~0.8$ after $300$ oscillation of $\Phi$, respectively.  
 \begin{figure}[t]
       \begin{flushleft}
  \begin{minipage}{\linewidth}
  \includegraphics[width=70mm]{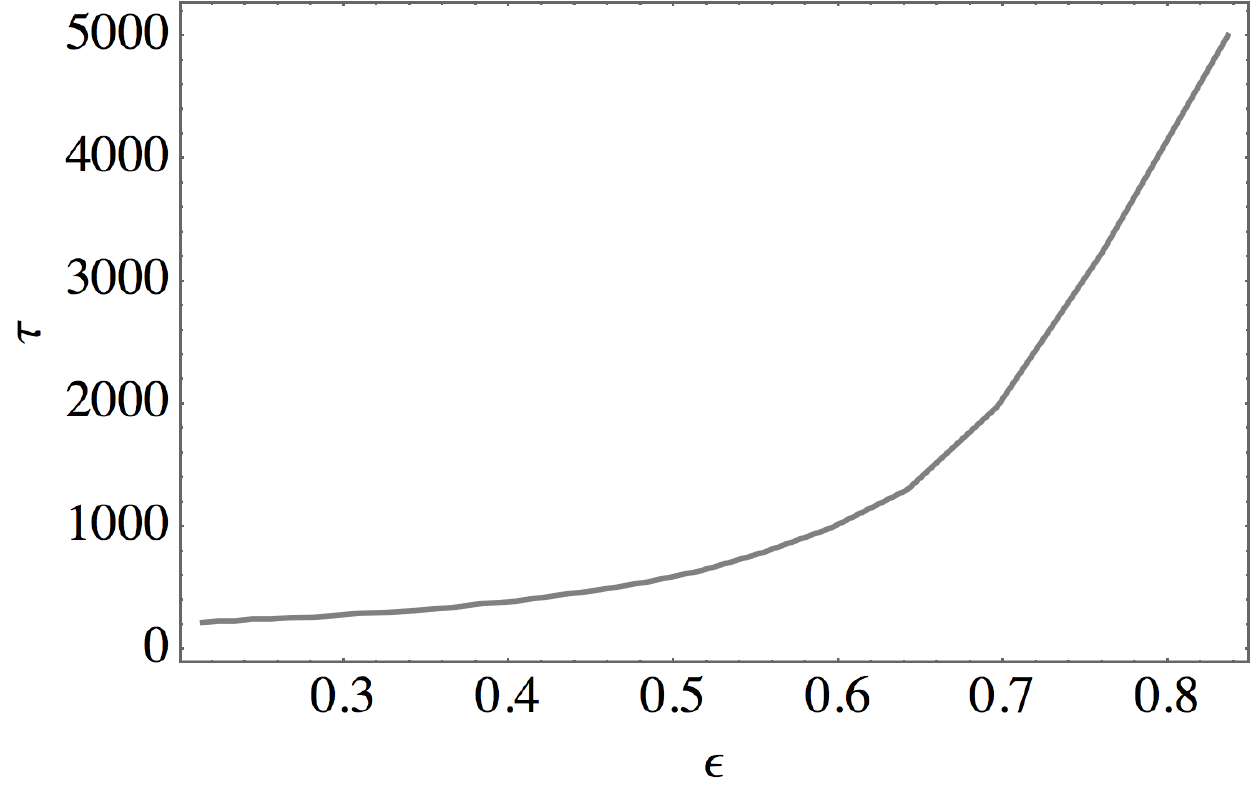}
 \end{minipage}
 \end{flushleft}
 \caption{\sl \small The relation between $\epsilon$ and the time when perturbations in some $k$ mode become the order $1$ in the extended Gaussian approximation.}
    \label{fig:fission}
\end{figure}
We also show in Fig.~\ref{fig:fission} the relation between the ellipticity and the time of $\delta \Phi=1$\footnote{The value itself is not important here because we are interested in relative behavior depending on $\epsilon$.} for some $k$ from the approximation ($\tau$).
The figures imply that $\delta\Phi$ has instability and grows exponentially in time, which is nothing but the broad resonance which is originated by the higher-order self-coupling of the complex scalar, and the perturbation growth is efficient for highly-elliptical case $\epsilon\ll1$. For a certain time scale, the fluctuation $\delta\Phi$ dominates over the background mode $\overline\Phi$ and the original Q-ball solution is no longer maintained. 

This time scale is expected to be related to the magnitude of the background oscillation: If the amplitude of the oscillation of $\overline\Phi$ is larger, the instability becomes stronger and the above time scale becomes shorter.
This means that highly-elliptical Q-ball has a short lifetime. This feature leads us to expect that the perturbation growth may be the seed of the fission of the elliptical Q-ball, since it is known that the fission is efficient for highly-elliptical case $\epsilon\ll1$~\cite{Hiramatsu:2010dx}.%
\footnote{Direct comparison of the transition time given by our approximation with that from the simulation results seems to be difficult because we need to extract information on the value of perturbation when the elliptical Q-ball forms. Therefore, we only compare the qualitative properties in this paper.}

\section{Conclusions and discussion}
\label{sec:7}
In this paper, we have studied the detailed properties of the elliptical Q-ball by the simulation and the semi-analysis. The elliptical Q-ball is the localized clump state with finite charge, which has the elliptical orbit in the complex field plane in Fig.~\ref{fig:nor}.%
\footnote{More precisely, the elliptical orbit is not closed, with its axes rotating.} This is different from the ground state Q-balls with the circular orbit. Those elliptical Q-balls correspond to the excited Q-ball which has larger energy per unit charge compared with the ground state Q-ball of the same charge. Although such an excited state of the Q-ball is discussed in many works of literature, the elliptical Q-ball formed in the numerical simulation is not described in detail. To investigate the detailed properties, we have performed the numerical lattice simulation in $3+1$ space-time dimension. We have first studied the spatial profiles of the elliptical Q-balls and showed that they have no nodes and some non-spherical Q-balls are formed. Furthermore, the (highly) elliptical orbit in the complex field plane shows that the elliptical Q-ball cannot be described by small perturbation against the ground state Q-ball.
We have also calculated the dispersion relation of the elliptical Q-balls, which depends on the ellipticity of the Q-balls as $E/Q\simeq (1+\epsilon^2)/2\epsilon$. We note that this relation is not well understood through the previous studies.
We have derived the approximate solution for the elliptical Q-balls that explains the numerical results, including the ellipticity dependence and the rotating property of the elliptical orbit, which leads to our conclusion that the elliptical Q-ball is an almost spherical~(nearly Gaussian) object that has the exciting energy depending on the ellipticity as $E/Q\simeq (1+\epsilon^2)/2\epsilon$, supported by the simulation and our approximation approaches.
We have also discussed the instability of the elliptical Q-ball against the small perturbation.

Several comments are in order. Throughout the paper, we have studied the elliptical Q-ball in gravity mediation. 
The gravity mediation model is useful since it is the simplest model in the sense that the properties including the Q-ball size and energy are given by the simple formulas. However, the existence and the properties of the elliptical Q-balls in the other models are non-trivial matter. For instance, in the gauge mediation models~\cite{Dvali:1997qv}, it has not been reported by the simulations whether the excited states are formed or not as transients. Furthermore, the ansatz for the Q-ball solution and its dispersion relation in the gauge mediation models are different from those in the gravity mediation models and they should be examined if the similar discussion as in this paper is valid.

There is also another type of Q-ball called New type Q-ball, which is realized when the both gravity and gauge mediation effects exist and gravity mediation dominates the potential. The formulas for the Q-ball properties are similar to the case when there exist only the gravity mediation effects, but the scales are different, e.g. soft mass and gravitino mass, which leads to the difference in the cosmological scenario (For instance, while the gravity mediation type Q-ball is unstable against the decay into the baryons, New-type Q-ball is stable and considered as a dark matter candidate).

In this work, we assumed the matter-dominated era during the Q-ball formation. Several scalar field baryogenesis mechanism~\cite{Sakstein:2017lfm,Sakstein:2017nns,Bettoni:2018utf,Liang:2019fkj}, however, occurs during the radiation dominated era. We expect that the elliptical Q-balls will be formed even during radiation dominated era (for generic potentials), since the reason that the elliptical Q-balls are formed as the excited transients is the energy gap between the initial coherent condensate and the final ground state Q-balls, which generally exists, especially independent of the way of cosmic expansion.

Let us also comment on the applications of our approximate theory to other non-topological solitons, $e.g.$ oscillon/I-ball~\cite{Bogolyubsky:1976yu,Gleiser:1993pt,Copeland:1995fq,Kasuya:2002zs,Amin:2010dc,Gleiser:2011xj,Amin:2011hj,Lozanov:2017hjm,Salmi:2012ta}  and boson star~\cite{Kaup:1968zz,Ruffini:1969qy}.  For instance, Refs.~\cite{Mukaida:2014oza,Eby:2018ufi}, suggest that oscillon/I-ball and boson star can be understood as Q-ball solutions associated with the particle number $U(1)$ charge, which might make it possible to derive some extended solutions of such non-topological solitons, using our extended theory of the Q-balls. These will also be explored in our forthcoming paper.


\vspace{+.4cm}

\section*{Acknowledgements}
J.P.H is supported by Korea NRF-2015R1A4A1042542. The work of F.H. and M.S. is supported in part by a JSPS Research Fellowship for Young Scientists Grant Number 17J07391 (F.H.) and 18J12023 (M.S.)

\bibliography{papers}

\begin{thebibliography}{38}%
\makeatletter
\providecommand \@ifxundefined [1]{%
 \@ifx{#1\undefined}
}%
\providecommand \@ifnum [1]{%
 \ifnum #1\expandafter \@firstoftwo
 \else \expandafter \@secondoftwo
 \fi
}%
\providecommand \@ifx [1]{%
 \ifx #1\expandafter \@firstoftwo
 \else \expandafter \@secondoftwo
 \fi
}%
\providecommand \natexlab [1]{#1}%
\providecommand \enquote  [1]{``#1''}%
\providecommand \bibnamefont  [1]{#1}%
\providecommand \bibfnamefont [1]{#1}%
\providecommand \citenamefont [1]{#1}%
\providecommand \href@noop [0]{\@secondoftwo}%
\providecommand \href [0]{\begingroup \@sanitize@url \@href}%
\providecommand \@href[1]{\@@startlink{#1}\@@href}%
\providecommand \@@href[1]{\endgroup#1\@@endlink}%
\providecommand \@sanitize@url [0]{\catcode `\\12\catcode `\$12\catcode
  `\&12\catcode `\#12\catcode `\^12\catcode `\_12\catcode `\%12\relax}%
\providecommand \@@startlink[1]{}%
\providecommand \@@endlink[0]{}%
\providecommand \url  [0]{\begingroup\@sanitize@url \@url }%
\providecommand \@url [1]{\endgroup\@href {#1}{\urlprefix }}%
\providecommand \urlprefix  [0]{URL }%
\providecommand \Eprint [0]{\href }%
\providecommand \doibase [0]{http://dx.doi.org/}%
\providecommand \selectlanguage [0]{\@gobble}%
\providecommand \bibinfo  [0]{\@secondoftwo}%
\providecommand \bibfield  [0]{\@secondoftwo}%
\providecommand \translation [1]{[#1]}%
\providecommand \BibitemOpen [0]{}%
\providecommand \bibitemStop [0]{}%
\providecommand \bibitemNoStop [0]{.\EOS\space}%
\providecommand \EOS [0]{\spacefactor3000\relax}%
\providecommand \BibitemShut  [1]{\csname bibitem#1\endcsname}%
\let\auto@bib@innerbib\@empty
\bibitem [{\citenamefont {Coleman}(1985)}]{Coleman:1985ki}%
  \BibitemOpen
  \bibfield  {author} {\bibinfo {author} {\bibfnamefont {S.~R.}\ \bibnamefont
  {Coleman}},\ }\href {\doibase 10.1016/0550-3213(85)90286-X,
  10.1016/0550-3213(86)90520-1} {\bibfield  {journal} {\bibinfo  {journal}
  {Nucl. Phys.}\ }\textbf {\bibinfo {volume} {B262}},\ \bibinfo {pages} {263}
  (\bibinfo {year} {1985})},\ \bibinfo {note} {[Erratum: Nucl.
  Phys.B269,744(1986)]}\BibitemShut {NoStop}%
\bibitem [{\citenamefont {Affleck}\ and\ \citenamefont
  {Dine}(1985)}]{Affleck:1984fy}%
  \BibitemOpen
  \bibfield  {author} {\bibinfo {author} {\bibfnamefont {I.}~\bibnamefont
  {Affleck}}\ and\ \bibinfo {author} {\bibfnamefont {M.}~\bibnamefont {Dine}},\
  }\href {\doibase 10.1016/0550-3213(85)90021-5} {\bibfield  {journal}
  {\bibinfo  {journal} {Nucl. Phys.}\ }\textbf {\bibinfo {volume} {B249}},\
  \bibinfo {pages} {361} (\bibinfo {year} {1985})}\BibitemShut {NoStop}%
\bibitem [{\citenamefont {Cohen}\ \emph {et~al.}(1986)\citenamefont {Cohen},
  \citenamefont {Coleman}, \citenamefont {Georgi},\ and\ \citenamefont
  {Manohar}}]{Cohen:1986ct}%
  \BibitemOpen
  \bibfield  {author} {\bibinfo {author} {\bibfnamefont {A.~G.}\ \bibnamefont
  {Cohen}}, \bibinfo {author} {\bibfnamefont {S.~R.}\ \bibnamefont {Coleman}},
  \bibinfo {author} {\bibfnamefont {H.}~\bibnamefont {Georgi}}, \ and\ \bibinfo
  {author} {\bibfnamefont {A.}~\bibnamefont {Manohar}},\ }\href {\doibase
  10.1016/0550-3213(86)90004-0} {\bibfield  {journal} {\bibinfo  {journal}
  {Nucl. Phys.}\ }\textbf {\bibinfo {volume} {B272}},\ \bibinfo {pages} {301}
  (\bibinfo {year} {1986})}\BibitemShut {NoStop}%
\bibitem [{\citenamefont {Dine}\ \emph {et~al.}(1996)\citenamefont {Dine},
  \citenamefont {Randall},\ and\ \citenamefont {Thomas}}]{Dine:1995kz}%
  \BibitemOpen
  \bibfield  {author} {\bibinfo {author} {\bibfnamefont {M.}~\bibnamefont
  {Dine}}, \bibinfo {author} {\bibfnamefont {L.}~\bibnamefont {Randall}}, \
  and\ \bibinfo {author} {\bibfnamefont {S.~D.}\ \bibnamefont {Thomas}},\
  }\href {\doibase 10.1016/0550-3213(95)00538-2} {\bibfield  {journal}
  {\bibinfo  {journal} {Nucl. Phys.}\ }\textbf {\bibinfo {volume} {B458}},\
  \bibinfo {pages} {291} (\bibinfo {year} {1996})},\ \Eprint
  {http://arxiv.org/abs/hep-ph/9507453} {arXiv:hep-ph/9507453 [hep-ph]}
  \BibitemShut {NoStop}%
\bibitem [{\citenamefont {Dvali}\ \emph {et~al.}(1998)\citenamefont {Dvali},
  \citenamefont {Kusenko},\ and\ \citenamefont {Shaposhnikov}}]{Dvali:1997qv}%
  \BibitemOpen
  \bibfield  {author} {\bibinfo {author} {\bibfnamefont {G.~R.}\ \bibnamefont
  {Dvali}}, \bibinfo {author} {\bibfnamefont {A.}~\bibnamefont {Kusenko}}, \
  and\ \bibinfo {author} {\bibfnamefont {M.~E.}\ \bibnamefont {Shaposhnikov}},\
  }\href {\doibase 10.1016/S0370-2693(97)01378-6} {\bibfield  {journal}
  {\bibinfo  {journal} {Phys. Lett.}\ }\textbf {\bibinfo {volume} {B417}},\
  \bibinfo {pages} {99} (\bibinfo {year} {1998})},\ \Eprint
  {http://arxiv.org/abs/hep-ph/9707423} {arXiv:hep-ph/9707423 [hep-ph]}
  \BibitemShut {NoStop}%
\bibitem [{\citenamefont {Kusenko}\ and\ \citenamefont
  {Shaposhnikov}(1998)}]{Kusenko:1997si}%
  \BibitemOpen
  \bibfield  {author} {\bibinfo {author} {\bibfnamefont {A.}~\bibnamefont
  {Kusenko}}\ and\ \bibinfo {author} {\bibfnamefont {M.~E.}\ \bibnamefont
  {Shaposhnikov}},\ }\href {\doibase 10.1016/S0370-2693(97)01375-0} {\bibfield
  {journal} {\bibinfo  {journal} {Phys. Lett.}\ }\textbf {\bibinfo {volume}
  {B418}},\ \bibinfo {pages} {46} (\bibinfo {year} {1998})},\ \Eprint
  {http://arxiv.org/abs/hep-ph/9709492} {arXiv:hep-ph/9709492 [hep-ph]}
  \BibitemShut {NoStop}%
\bibitem [{\citenamefont {Enqvist}\ and\ \citenamefont
  {McDonald}(1998{\natexlab{a}})}]{Enqvist:1997si}%
  \BibitemOpen
  \bibfield  {author} {\bibinfo {author} {\bibfnamefont {K.}~\bibnamefont
  {Enqvist}}\ and\ \bibinfo {author} {\bibfnamefont {J.}~\bibnamefont
  {McDonald}},\ }\href {\doibase 10.1016/S0370-2693(98)00271-8} {\bibfield
  {journal} {\bibinfo  {journal} {Phys. Lett.}\ }\textbf {\bibinfo {volume}
  {B425}},\ \bibinfo {pages} {309} (\bibinfo {year} {1998}{\natexlab{a}})},\
  \Eprint {http://arxiv.org/abs/hep-ph/9711514} {arXiv:hep-ph/9711514 [hep-ph]}
  \BibitemShut {NoStop}%
\bibitem [{\citenamefont {Enqvist}\ and\ \citenamefont
  {McDonald}(1998{\natexlab{b}})}]{Enqvist:1998xd}%
  \BibitemOpen
  \bibfield  {author} {\bibinfo {author} {\bibfnamefont {K.}~\bibnamefont
  {Enqvist}}\ and\ \bibinfo {author} {\bibfnamefont {J.}~\bibnamefont
  {McDonald}},\ }\href {\doibase 10.1016/S0370-2693(98)01078-8} {\bibfield
  {journal} {\bibinfo  {journal} {Phys. Lett.}\ }\textbf {\bibinfo {volume}
  {B440}},\ \bibinfo {pages} {59} (\bibinfo {year} {1998}{\natexlab{b}})},\
  \Eprint {http://arxiv.org/abs/hep-ph/9807269} {arXiv:hep-ph/9807269 [hep-ph]}
  \BibitemShut {NoStop}%
\bibitem [{\citenamefont {Enqvist}\ and\ \citenamefont
  {McDonald}(1999)}]{Enqvist:1998en}%
  \BibitemOpen
  \bibfield  {author} {\bibinfo {author} {\bibfnamefont {K.}~\bibnamefont
  {Enqvist}}\ and\ \bibinfo {author} {\bibfnamefont {J.}~\bibnamefont
  {McDonald}},\ }\href {\doibase 10.1016/S0550-3213(98)00695-6} {\bibfield
  {journal} {\bibinfo  {journal} {Nucl. Phys.}\ }\textbf {\bibinfo {volume}
  {B538}},\ \bibinfo {pages} {321} (\bibinfo {year} {1999})},\ \Eprint
  {http://arxiv.org/abs/hep-ph/9803380} {arXiv:hep-ph/9803380 [hep-ph]}
  \BibitemShut {NoStop}%
\bibitem [{\citenamefont {Kasuya}\ and\ \citenamefont
  {Kawasaki}(2000{\natexlab{a}})}]{Kasuya:1999wu}%
  \BibitemOpen
  \bibfield  {author} {\bibinfo {author} {\bibfnamefont {S.}~\bibnamefont
  {Kasuya}}\ and\ \bibinfo {author} {\bibfnamefont {M.}~\bibnamefont
  {Kawasaki}},\ }\href {\doibase 10.1103/PhysRevD.61.041301} {\bibfield
  {journal} {\bibinfo  {journal} {Phys. Rev.}\ }\textbf {\bibinfo {volume}
  {D61}},\ \bibinfo {pages} {041301} (\bibinfo {year} {2000}{\natexlab{a}})},\
  \Eprint {http://arxiv.org/abs/hep-ph/9909509} {arXiv:hep-ph/9909509 [hep-ph]}
  \BibitemShut {NoStop}%
\bibitem [{\citenamefont {Kasuya}\ and\ \citenamefont
  {Kawasaki}(2000{\natexlab{b}})}]{Kasuya:2000wx}%
  \BibitemOpen
  \bibfield  {author} {\bibinfo {author} {\bibfnamefont {S.}~\bibnamefont
  {Kasuya}}\ and\ \bibinfo {author} {\bibfnamefont {M.}~\bibnamefont
  {Kawasaki}},\ }\href {\doibase 10.1103/PhysRevD.62.023512} {\bibfield
  {journal} {\bibinfo  {journal} {Phys. Rev.}\ }\textbf {\bibinfo {volume}
  {D62}},\ \bibinfo {pages} {023512} (\bibinfo {year} {2000}{\natexlab{b}})},\
  \Eprint {http://arxiv.org/abs/hep-ph/0002285} {arXiv:hep-ph/0002285 [hep-ph]}
  \BibitemShut {NoStop}%
\bibitem [{\citenamefont {Enqvist}\ \emph {et~al.}(2001)\citenamefont
  {Enqvist}, \citenamefont {Jokinen}, \citenamefont {Multamaki},\ and\
  \citenamefont {Vilja}}]{Enqvist:2000cq}%
  \BibitemOpen
  \bibfield  {author} {\bibinfo {author} {\bibfnamefont {K.}~\bibnamefont
  {Enqvist}}, \bibinfo {author} {\bibfnamefont {A.}~\bibnamefont {Jokinen}},
  \bibinfo {author} {\bibfnamefont {T.}~\bibnamefont {Multamaki}}, \ and\
  \bibinfo {author} {\bibfnamefont {I.}~\bibnamefont {Vilja}},\ }\href
  {\doibase 10.1103/PhysRevD.63.083501} {\bibfield  {journal} {\bibinfo
  {journal} {Phys. Rev.}\ }\textbf {\bibinfo {volume} {D63}},\ \bibinfo {pages}
  {083501} (\bibinfo {year} {2001})},\ \Eprint
  {http://arxiv.org/abs/hep-ph/0011134} {arXiv:hep-ph/0011134 [hep-ph]}
  \BibitemShut {NoStop}%
\bibitem [{\citenamefont {Kasuya}\ and\ \citenamefont
  {Kawasaki}(2001)}]{Kasuya:2001hg}%
  \BibitemOpen
  \bibfield  {author} {\bibinfo {author} {\bibfnamefont {S.}~\bibnamefont
  {Kasuya}}\ and\ \bibinfo {author} {\bibfnamefont {M.}~\bibnamefont
  {Kawasaki}},\ }\href {\doibase 10.1103/PhysRevD.64.123515} {\bibfield
  {journal} {\bibinfo  {journal} {Phys. Rev.}\ }\textbf {\bibinfo {volume}
  {D64}},\ \bibinfo {pages} {123515} (\bibinfo {year} {2001})},\ \Eprint
  {http://arxiv.org/abs/hep-ph/0106119} {arXiv:hep-ph/0106119 [hep-ph]}
  \BibitemShut {NoStop}%
\bibitem [{\citenamefont {Hiramatsu}\ \emph {et~al.}(2010)\citenamefont
  {Hiramatsu}, \citenamefont {Kawasaki},\ and\ \citenamefont
  {Takahashi}}]{Hiramatsu:2010dx}%
  \BibitemOpen
  \bibfield  {author} {\bibinfo {author} {\bibfnamefont {T.}~\bibnamefont
  {Hiramatsu}}, \bibinfo {author} {\bibfnamefont {M.}~\bibnamefont {Kawasaki}},
  \ and\ \bibinfo {author} {\bibfnamefont {F.}~\bibnamefont {Takahashi}},\
  }\href {\doibase 10.1088/1475-7516/2010/06/008} {\bibfield  {journal}
  {\bibinfo  {journal} {JCAP}\ }\textbf {\bibinfo {volume} {1006}},\ \bibinfo
  {pages} {008} (\bibinfo {year} {2010})},\ \Eprint
  {http://arxiv.org/abs/1003.1779} {arXiv:1003.1779 [hep-ph]} \BibitemShut
  {NoStop}%
\bibitem [{\citenamefont {Enqvist}\ and\ \citenamefont
  {McDonald}(2000)}]{Enqvist:1999mv}%
  \BibitemOpen
  \bibfield  {author} {\bibinfo {author} {\bibfnamefont {K.}~\bibnamefont
  {Enqvist}}\ and\ \bibinfo {author} {\bibfnamefont {J.}~\bibnamefont
  {McDonald}},\ }\href {\doibase 10.1016/S0550-3213(99)00776-2,
  10.1016/S0550-3213(00)00304-7} {\bibfield  {journal} {\bibinfo  {journal}
  {Nucl. Phys.}\ }\textbf {\bibinfo {volume} {B570}},\ \bibinfo {pages} {407}
  (\bibinfo {year} {2000})},\ \bibinfo {note} {[Erratum: Nucl.
  Phys.B582,763(2000)]},\ \Eprint {http://arxiv.org/abs/hep-ph/9908316}
  {arXiv:hep-ph/9908316 [hep-ph]} \BibitemShut {NoStop}%
\bibitem [{\citenamefont {Lozanov}\ and\ \citenamefont
  {Amin}(2014)}]{Lozanov:2014zfa}%
  \BibitemOpen
  \bibfield  {author} {\bibinfo {author} {\bibfnamefont {K.~D.}\ \bibnamefont
  {Lozanov}}\ and\ \bibinfo {author} {\bibfnamefont {M.~A.}\ \bibnamefont
  {Amin}},\ }\href {\doibase 10.1103/PhysRevD.90.083528} {\bibfield  {journal}
  {\bibinfo  {journal} {Phys. Rev.}\ }\textbf {\bibinfo {volume} {D90}},\
  \bibinfo {pages} {083528} (\bibinfo {year} {2014})},\ \Eprint
  {http://arxiv.org/abs/1408.1811} {arXiv:1408.1811 [hep-ph]} \BibitemShut
  {NoStop}%
\bibitem [{\citenamefont {Volkov}\ and\ \citenamefont
  {Wohnert}(2002)}]{Volkov:2002aj}%
  \BibitemOpen
  \bibfield  {author} {\bibinfo {author} {\bibfnamefont {M.~S.}\ \bibnamefont
  {Volkov}}\ and\ \bibinfo {author} {\bibfnamefont {E.}~\bibnamefont
  {Wohnert}},\ }\href {\doibase 10.1103/PhysRevD.66.085003} {\bibfield
  {journal} {\bibinfo  {journal} {Phys. Rev.}\ }\textbf {\bibinfo {volume}
  {D66}},\ \bibinfo {pages} {085003} (\bibinfo {year} {2002})},\ \Eprint
  {http://arxiv.org/abs/hep-th/0205157} {arXiv:hep-th/0205157 [hep-th]}
  \BibitemShut {NoStop}%
\bibitem [{\citenamefont {Polyakov}\ and\ \citenamefont
  {Schweitzer}(2018)}]{Polyakov:2018zvc}%
  \BibitemOpen
  \bibfield  {author} {\bibinfo {author} {\bibfnamefont {M.~V.}\ \bibnamefont
  {Polyakov}}\ and\ \bibinfo {author} {\bibfnamefont {P.}~\bibnamefont
  {Schweitzer}},\ }\href {\doibase 10.1142/S0217751X18300259} {\bibfield
  {journal} {\bibinfo  {journal} {Int. J. Mod. Phys.}\ }\textbf {\bibinfo
  {volume} {A33}},\ \bibinfo {pages} {1830025} (\bibinfo {year} {2018})},\
  \Eprint {http://arxiv.org/abs/1805.06596} {arXiv:1805.06596 [hep-ph]}
  \BibitemShut {NoStop}%
\bibitem [{\citenamefont {Lee}\ and\ \citenamefont {Pang}(1992)}]{Lee:1991ax}%
  \BibitemOpen
  \bibfield  {author} {\bibinfo {author} {\bibfnamefont {T.~D.}\ \bibnamefont
  {Lee}}\ and\ \bibinfo {author} {\bibfnamefont {Y.}~\bibnamefont {Pang}},\
  }\href {\doibase 10.1016/0370-1573(92)90064-7} {\bibfield  {journal}
  {\bibinfo  {journal} {Phys. Rept.}\ }\textbf {\bibinfo {volume} {221}},\
  \bibinfo {pages} {251} (\bibinfo {year} {1992})},\ \bibinfo {note}
  {[,169(1991)]}\BibitemShut {NoStop}%
\bibitem [{\citenamefont {Smolyakov}(2018)}]{Smolyakov:2017axd}%
  \BibitemOpen
  \bibfield  {author} {\bibinfo {author} {\bibfnamefont {M.~N.}\ \bibnamefont
  {Smolyakov}},\ }\href {\doibase 10.1103/PhysRevD.97.045011} {\bibfield
  {journal} {\bibinfo  {journal} {Phys. Rev.}\ }\textbf {\bibinfo {volume}
  {D97}},\ \bibinfo {pages} {045011} (\bibinfo {year} {2018})},\ \Eprint
  {http://arxiv.org/abs/1711.05730} {arXiv:1711.05730 [hep-th]} \BibitemShut
  {NoStop}%
\bibitem [{\citenamefont {Sainio}(2012)}]{Sainio:2012mw}%
  \BibitemOpen
  \bibfield  {author} {\bibinfo {author} {\bibfnamefont {J.}~\bibnamefont
  {Sainio}},\ }\href {\doibase 10.1088/1475-7516/2012/04/038} {\bibfield
  {journal} {\bibinfo  {journal} {JCAP}\ }\textbf {\bibinfo {volume} {1204}},\
  \bibinfo {pages} {038} (\bibinfo {year} {2012})},\ \Eprint
  {http://arxiv.org/abs/1201.5029} {arXiv:1201.5029 [astro-ph.IM]} \BibitemShut
  {NoStop}%
\bibitem [{\citenamefont {Bogolyubsky}\ and\ \citenamefont
  {Makhankov}(1976)}]{Bogolyubsky:1976yu}%
  \BibitemOpen
  \bibfield  {author} {\bibinfo {author} {\bibfnamefont {I.~L.}\ \bibnamefont
  {Bogolyubsky}}\ and\ \bibinfo {author} {\bibfnamefont {V.~G.}\ \bibnamefont
  {Makhankov}},\ }\href@noop {} {\bibfield  {journal} {\bibinfo  {journal}
  {Pisma Zh. Eksp. Teor. Fiz.}\ }\textbf {\bibinfo {volume} {24}},\ \bibinfo
  {pages} {15} (\bibinfo {year} {1976})}\BibitemShut {NoStop}%
\bibitem [{\citenamefont {Gleiser}(1994)}]{Gleiser:1993pt}%
  \BibitemOpen
  \bibfield  {author} {\bibinfo {author} {\bibfnamefont {M.}~\bibnamefont
  {Gleiser}},\ }\href {\doibase 10.1103/PhysRevD.49.2978} {\bibfield  {journal}
  {\bibinfo  {journal} {Phys. Rev.}\ }\textbf {\bibinfo {volume} {D49}},\
  \bibinfo {pages} {2978} (\bibinfo {year} {1994})},\ \Eprint
  {http://arxiv.org/abs/hep-ph/9308279} {arXiv:hep-ph/9308279 [hep-ph]}
  \BibitemShut {NoStop}%
\bibitem [{\citenamefont {Kasuya}\ \emph {et~al.}(2003)\citenamefont {Kasuya},
  \citenamefont {Kawasaki},\ and\ \citenamefont {Takahashi}}]{Kasuya:2002zs}%
  \BibitemOpen
  \bibfield  {author} {\bibinfo {author} {\bibfnamefont {S.}~\bibnamefont
  {Kasuya}}, \bibinfo {author} {\bibfnamefont {M.}~\bibnamefont {Kawasaki}}, \
  and\ \bibinfo {author} {\bibfnamefont {F.}~\bibnamefont {Takahashi}},\ }\href
  {\doibase 10.1016/S0370-2693(03)00344-7} {\bibfield  {journal} {\bibinfo
  {journal} {Phys. Lett.}\ }\textbf {\bibinfo {volume} {B559}},\ \bibinfo
  {pages} {99} (\bibinfo {year} {2003})},\ \Eprint
  {http://arxiv.org/abs/hep-ph/0209358} {arXiv:hep-ph/0209358 [hep-ph]}
  \BibitemShut {NoStop}%
\bibitem [{\citenamefont {Sakstein}\ and\ \citenamefont
  {Trodden}(2017)}]{Sakstein:2017lfm}%
  \BibitemOpen
  \bibfield  {author} {\bibinfo {author} {\bibfnamefont {J.}~\bibnamefont
  {Sakstein}}\ and\ \bibinfo {author} {\bibfnamefont {M.}~\bibnamefont
  {Trodden}},\ }\href {\doibase 10.1016/j.physletb.2017.09.059} {\bibfield
  {journal} {\bibinfo  {journal} {Phys. Lett.}\ }\textbf {\bibinfo {volume}
  {B774}},\ \bibinfo {pages} {183} (\bibinfo {year} {2017})},\ \Eprint
  {http://arxiv.org/abs/1703.10103} {arXiv:1703.10103 [hep-ph]} \BibitemShut
  {NoStop}%
\bibitem [{\citenamefont {Sakstein}\ and\ \citenamefont
  {Solomon}(2017)}]{Sakstein:2017nns}%
  \BibitemOpen
  \bibfield  {author} {\bibinfo {author} {\bibfnamefont {J.}~\bibnamefont
  {Sakstein}}\ and\ \bibinfo {author} {\bibfnamefont {A.~R.}\ \bibnamefont
  {Solomon}},\ }\href {\doibase 10.1016/j.physletb.2017.08.039} {\bibfield
  {journal} {\bibinfo  {journal} {Phys. Lett.}\ }\textbf {\bibinfo {volume}
  {B773}},\ \bibinfo {pages} {186} (\bibinfo {year} {2017})},\ \Eprint
  {http://arxiv.org/abs/1705.10695} {arXiv:1705.10695 [hep-ph]} \BibitemShut
  {NoStop}%
\bibitem [{\citenamefont {Bettoni}\ and\ \citenamefont
  {Rubio}(2018)}]{Bettoni:2018utf}%
  \BibitemOpen
  \bibfield  {author} {\bibinfo {author} {\bibfnamefont {D.}~\bibnamefont
  {Bettoni}}\ and\ \bibinfo {author} {\bibfnamefont {J.}~\bibnamefont
  {Rubio}},\ }\href {\doibase 10.1016/j.physletb.2018.07.046} {\bibfield
  {journal} {\bibinfo  {journal} {Phys. Lett.}\ }\textbf {\bibinfo {volume}
  {B784}},\ \bibinfo {pages} {122} (\bibinfo {year} {2018})},\ \Eprint
  {http://arxiv.org/abs/1805.02669} {arXiv:1805.02669 [astro-ph.CO]}
  \BibitemShut {NoStop}%
\bibitem [{\citenamefont {Liang}\ \emph {et~al.}(2019)\citenamefont {Liang},
  \citenamefont {Sakstein},\ and\ \citenamefont {Trodden}}]{Liang:2019fkj}%
  \BibitemOpen
  \bibfield  {author} {\bibinfo {author} {\bibfnamefont {Q.}~\bibnamefont
  {Liang}}, \bibinfo {author} {\bibfnamefont {J.}~\bibnamefont {Sakstein}}, \
  and\ \bibinfo {author} {\bibfnamefont {M.}~\bibnamefont {Trodden}},\
  }\href@noop {} {\  (\bibinfo {year} {2019})},\ \Eprint
  {http://arxiv.org/abs/1904.10510} {arXiv:1904.10510 [hep-ph]} \BibitemShut
  {NoStop}%
\bibitem [{\citenamefont {Copeland}\ \emph {et~al.}(1995)\citenamefont
  {Copeland}, \citenamefont {Gleiser},\ and\ \citenamefont
  {Muller}}]{Copeland:1995fq}%
  \BibitemOpen
  \bibfield  {author} {\bibinfo {author} {\bibfnamefont {E.~J.}\ \bibnamefont
  {Copeland}}, \bibinfo {author} {\bibfnamefont {M.}~\bibnamefont {Gleiser}}, \
  and\ \bibinfo {author} {\bibfnamefont {H.~R.}\ \bibnamefont {Muller}},\
  }\href {\doibase 10.1103/PhysRevD.52.1920} {\bibfield  {journal} {\bibinfo
  {journal} {Phys. Rev.}\ }\textbf {\bibinfo {volume} {D52}},\ \bibinfo {pages}
  {1920} (\bibinfo {year} {1995})},\ \Eprint
  {http://arxiv.org/abs/hep-ph/9503217} {arXiv:hep-ph/9503217 [hep-ph]}
  \BibitemShut {NoStop}%
\bibitem [{\citenamefont {Amin}\ \emph {et~al.}(2010)\citenamefont {Amin},
  \citenamefont {Easther},\ and\ \citenamefont {Finkel}}]{Amin:2010dc}%
  \BibitemOpen
  \bibfield  {author} {\bibinfo {author} {\bibfnamefont {M.~A.}\ \bibnamefont
  {Amin}}, \bibinfo {author} {\bibfnamefont {R.}~\bibnamefont {Easther}}, \
  and\ \bibinfo {author} {\bibfnamefont {H.}~\bibnamefont {Finkel}},\ }\href
  {\doibase 10.1088/1475-7516/2010/12/001} {\bibfield  {journal} {\bibinfo
  {journal} {JCAP}\ }\textbf {\bibinfo {volume} {1012}},\ \bibinfo {pages}
  {001} (\bibinfo {year} {2010})},\ \Eprint {http://arxiv.org/abs/1009.2505}
  {arXiv:1009.2505 [astro-ph.CO]} \BibitemShut {NoStop}%
\bibitem [{\citenamefont {Gleiser}\ \emph {et~al.}(2011)\citenamefont
  {Gleiser}, \citenamefont {Graham},\ and\ \citenamefont
  {Stamatopoulos}}]{Gleiser:2011xj}%
  \BibitemOpen
  \bibfield  {author} {\bibinfo {author} {\bibfnamefont {M.}~\bibnamefont
  {Gleiser}}, \bibinfo {author} {\bibfnamefont {N.}~\bibnamefont {Graham}}, \
  and\ \bibinfo {author} {\bibfnamefont {N.}~\bibnamefont {Stamatopoulos}},\
  }\href {\doibase 10.1103/PhysRevD.83.096010} {\bibfield  {journal} {\bibinfo
  {journal} {Phys. Rev.}\ }\textbf {\bibinfo {volume} {D83}},\ \bibinfo {pages}
  {096010} (\bibinfo {year} {2011})},\ \Eprint {http://arxiv.org/abs/1103.1911}
  {arXiv:1103.1911 [hep-th]} \BibitemShut {NoStop}%
\bibitem [{\citenamefont {Amin}\ \emph {et~al.}(2012)\citenamefont {Amin},
  \citenamefont {Easther}, \citenamefont {Finkel}, \citenamefont {Flauger},\
  and\ \citenamefont {Hertzberg}}]{Amin:2011hj}%
  \BibitemOpen
  \bibfield  {author} {\bibinfo {author} {\bibfnamefont {M.~A.}\ \bibnamefont
  {Amin}}, \bibinfo {author} {\bibfnamefont {R.}~\bibnamefont {Easther}},
  \bibinfo {author} {\bibfnamefont {H.}~\bibnamefont {Finkel}}, \bibinfo
  {author} {\bibfnamefont {R.}~\bibnamefont {Flauger}}, \ and\ \bibinfo
  {author} {\bibfnamefont {M.~P.}\ \bibnamefont {Hertzberg}},\ }\href {\doibase
  10.1103/PhysRevLett.108.241302} {\bibfield  {journal} {\bibinfo  {journal}
  {Phys. Rev. Lett.}\ }\textbf {\bibinfo {volume} {108}},\ \bibinfo {pages}
  {241302} (\bibinfo {year} {2012})},\ \Eprint {http://arxiv.org/abs/1106.3335}
  {arXiv:1106.3335 [astro-ph.CO]} \BibitemShut {NoStop}%
\bibitem [{\citenamefont {Lozanov}\ and\ \citenamefont
  {Amin}(2018)}]{Lozanov:2017hjm}%
  \BibitemOpen
  \bibfield  {author} {\bibinfo {author} {\bibfnamefont {K.~D.}\ \bibnamefont
  {Lozanov}}\ and\ \bibinfo {author} {\bibfnamefont {M.~A.}\ \bibnamefont
  {Amin}},\ }\href {\doibase 10.1103/PhysRevD.97.023533} {\bibfield  {journal}
  {\bibinfo  {journal} {Phys. Rev.}\ }\textbf {\bibinfo {volume} {D97}},\
  \bibinfo {pages} {023533} (\bibinfo {year} {2018})},\ \Eprint
  {http://arxiv.org/abs/1710.06851} {arXiv:1710.06851 [astro-ph.CO]}
  \BibitemShut {NoStop}%
\bibitem [{\citenamefont {Salmi}\ and\ \citenamefont
  {Hindmarsh}(2012)}]{Salmi:2012ta}%
  \BibitemOpen
  \bibfield  {author} {\bibinfo {author} {\bibfnamefont {P.}~\bibnamefont
  {Salmi}}\ and\ \bibinfo {author} {\bibfnamefont {M.}~\bibnamefont
  {Hindmarsh}},\ }\href {\doibase 10.1103/PhysRevD.85.085033} {\bibfield
  {journal} {\bibinfo  {journal} {Phys. Rev.}\ }\textbf {\bibinfo {volume}
  {D85}},\ \bibinfo {pages} {085033} (\bibinfo {year} {2012})},\ \Eprint
  {http://arxiv.org/abs/1201.1934} {arXiv:1201.1934 [hep-th]} \BibitemShut
  {NoStop}%
\bibitem [{\citenamefont {Kaup}(1968)}]{Kaup:1968zz}%
  \BibitemOpen
  \bibfield  {author} {\bibinfo {author} {\bibfnamefont {D.~J.}\ \bibnamefont
  {Kaup}},\ }\href {\doibase 10.1103/PhysRev.172.1331} {\bibfield  {journal}
  {\bibinfo  {journal} {Phys. Rev.}\ }\textbf {\bibinfo {volume} {172}},\
  \bibinfo {pages} {1331} (\bibinfo {year} {1968})}\BibitemShut {NoStop}%
\bibitem [{\citenamefont {Ruffini}\ and\ \citenamefont
  {Bonazzola}(1969)}]{Ruffini:1969qy}%
  \BibitemOpen
  \bibfield  {author} {\bibinfo {author} {\bibfnamefont {R.}~\bibnamefont
  {Ruffini}}\ and\ \bibinfo {author} {\bibfnamefont {S.}~\bibnamefont
  {Bonazzola}},\ }\href {\doibase 10.1103/PhysRev.187.1767} {\bibfield
  {journal} {\bibinfo  {journal} {Phys. Rev.}\ }\textbf {\bibinfo {volume}
  {187}},\ \bibinfo {pages} {1767} (\bibinfo {year} {1969})}\BibitemShut
  {NoStop}%
\bibitem [{\citenamefont {Mukaida}\ and\ \citenamefont
  {Takimoto}(2014)}]{Mukaida:2014oza}%
  \BibitemOpen
  \bibfield  {author} {\bibinfo {author} {\bibfnamefont {K.}~\bibnamefont
  {Mukaida}}\ and\ \bibinfo {author} {\bibfnamefont {M.}~\bibnamefont
  {Takimoto}},\ }\href {\doibase 10.1088/1475-7516/2014/08/051} {\bibfield
  {journal} {\bibinfo  {journal} {JCAP}\ }\textbf {\bibinfo {volume} {1408}},\
  \bibinfo {pages} {051} (\bibinfo {year} {2014})},\ \Eprint
  {http://arxiv.org/abs/1405.3233} {arXiv:1405.3233 [hep-ph]} \BibitemShut
  {NoStop}%
\bibitem [{\citenamefont {Eby}\ \emph {et~al.}(2018)\citenamefont {Eby},
  \citenamefont {Mukaida}, \citenamefont {Takimoto}, \citenamefont
  {Wijewardhana},\ and\ \citenamefont {Yamada}}]{Eby:2018ufi}%
  \BibitemOpen
  \bibfield  {author} {\bibinfo {author} {\bibfnamefont {J.}~\bibnamefont
  {Eby}}, \bibinfo {author} {\bibfnamefont {K.}~\bibnamefont {Mukaida}},
  \bibinfo {author} {\bibfnamefont {M.}~\bibnamefont {Takimoto}}, \bibinfo
  {author} {\bibfnamefont {L.~C.~R.}\ \bibnamefont {Wijewardhana}}, \ and\
  \bibinfo {author} {\bibfnamefont {M.}~\bibnamefont {Yamada}},\ }\href@noop {}
  {\  (\bibinfo {year} {2018})},\ \Eprint {http://arxiv.org/abs/1807.09795}
  {arXiv:1807.09795 [hep-ph]} \BibitemShut {NoStop}%
\end{thebibliography}%
\end{document}